\documentclass[aps,pra,amsmath,amssymb,twocolumn,preprintnumbers,superscriptaddress,10pt,longbibliography, nofootinbib]{revtex4-2}
\usepackage[utf8]{inputenc}                                             % Input encoding

\usepackage[dvipsnames]{xcolor}                                                     % Colors
\usepackage{times}

\usepackage{amsmath,amssymb}                                            % Math
\usepackage{amsthm}                                                     % Theorem environments
\usepackage[italicdiff]{physics}                                        % Physics
\usepackage{bm}                                                         % Bold math
\usepackage{tensor}                                                     % Tensors
\usepackage{graphicx}                                                   % Include figure files
\usepackage{placeins}

\usepackage{hyperref}                                                   % Hyperlinks

\newcommand*{\reals}{\mathbb{R}}                                        % Set of real numbers
\renewcommand*{\ket}[1]{\left| #1 \right\rangle}                        % Ket
\renewcommand*{\bra}[1]{\left\langle #1 \right|}                        % Bra
\newcommand*{\nth}{n_\text{th.}}

\newcommand*{\fref}[1]{Fig.~\ref{#1}}                                   % Figure reference
\newcommand*{\tref}[1]{Tab.~\ref{#1}}                                   % Table reference
\begin{document}

%===== Preprint ===============
%\preprint{}

%===== Title ==================
\title{Witnesses of Genuine Multipartite Entanglement and Nonlocal Measurement Back-action for Raman-scattering Quantum Systems}

%===== Authors ================
\author{Kai Ryen Bush}
\email{Kai.Ryen@usn.no}
\affiliation{Department of Science and Industry Systems, University of South-Eastern Norway, 3616 Kongsberg, Norway}

\author{Kjetil B{\o}rkje}
\email{Kjetil.Borkje@usn.no}
\affiliation{Department of Science and Industry Systems, University of South-Eastern Norway, 3616 Kongsberg, Norway}

%===== Date ===================
\date{\today}

%===== Abstract ===============
\begin{abstract}
Entanglement between remote quantum mechanical systems enables a range of quantum information tasks in communication, computation and distributed sensing. Large numbers of entangled subsystems also require experimentally accessible and practically feasible methods of verifying the genuine, i.e., simultaneous, entanglement of all subsystems. We have derived a class of entanglement witnesses suitable for $W$-states, which are states where a single excitation is coherently distributed across subsystems initially in their ground state or a state with low thermal occupation, e.g., via detection of a Raman-scattered photon. The entanglement is witnessed through violation of an inequality involving number statistics, which can be measured via detection of subsequent Raman-scattered photons. Unlike conventional, partially tomographic, witnesses, our method is experimentally accessible for both multipartite and continuous variable systems. The thermal robustness of the method is quantified by the initial thermal occupations for which violation occurs. As an alternative approach, we have derived an inequality which tests the nonlocal, or quantum coherent, nature of the photon measurement backaction which produces the $W$-state. Violation of this alternative inequality implies the entanglement of the resulting state given the assumption of a separable initial state, under less stringent thermal constraints than the general entanglement witness. Our results are applicable to all Raman-scattering systems which can exhibit sufficient degrees of quantum indistinguishability.
\end{abstract}

%===== Keywords ===============
%% Use showkeys class option if keyword display desired
%\keywords{}

%===== Maketitle ==============
\maketitle

\section{Introduction}
Entanglement of remote quantum systems is both of technological and fundamental interest. Bipartite entanglement of two distant systems can, for example, be exploited to teleport quantum information over long distances \cite{Bennett1993PRL}. Multipartite entanglement between many modes can provide further improvement to quantum communication protocols \cite{kruszynska2006PRA} or bring quantum advantage to distributed sensing schemes \cite{zhang2021QST}.

Quantum systems whose states can transition by Raman-scattering of photons are particularly suited for remote entanglement. A well-known example is the entanglement generation involved in the DLCZ protocol \cite{Duan2001Nature}. Here, starting from the ground state, the detection of a single photon scattered due to a Stokes transition projects two atomic ensembles into a Bell state $\propto \ket{1,0} + \ket{0, 1}$. This is caused by a loss of which-path information regarding the photon's origin after passing through a beam splitter. This procedure was first experimentally realized in Ref. \cite{Chou2005Nature}. Similar experiments have later been implemented to create heralded entanglement between intrinsic material resonances in diamond \cite{Lee2011Science}, between ions embedded in solid-state crystals \cite{Usmani2012NatPhys}, and between engineered micromechanical oscillators \cite{riedinger2018Nature}.

The detection of Raman-scattered photons can by the same principle also herald multipartite entanglement between a large number of subsystems. This can be achieved by having the scattered optical fields from the subsystems pass through a linear optical network consisting of beam splitters, mirrors, and phase shifters, before reaching a photodetector. The resulting entangled state for the $N$-partite case, if starting from the ground state, is the generalized $W$-state 
\begin{equation}
    \ket{W_N}=\frac{1}{\sqrt{N}}(\ket{1,0,0,\ldots}+\ket{0,1,0,\ldots}+\ket{0,0,1,\ldots}+\ldots),
\end{equation}
where all $N$ subsystems are entangled. $W$-states have the notable feature that when removing (or, mathematically, tracing over) one of the subsystems, the other $N-1$ subsystems remain in an entangled state. This robustness has potential value for quantum communication protocols \cite{Joo2002, Joo2003NJP, Agrawal2006PRA, Liu2012IJTP, li2024quantum}. Multipartite $W$-states have previously been realized in optical experiments involving entanglement between polarization modes (so-called photonic qbits) \cite{zou2002generation, kiesel2003three, tashima2009local},  frequencies \cite{Fang2019PRL} and photon numbers \cite{grafe2014chip}.

Once an entangled state has been produced, the next experimental challenge is how to verify the presence of entanglement. One strategy is to characterize the state itself by quantum state tomography. However, this generally requires a large number of measurements, and may or may not be accessible in a given experimental setup. For an increasing number of entangled subsystems, and in particular for continuous variable systems, this quickly becomes impractical. Another option is to construct an \emph{entanglement witness} \cite{Horodecki2009RMP, guhne2009PR, Chruscinski2014IOP, HilleryZubairy2006PRL, Borkje2011PRL, Guhne2010NJP} based on observable expectation values, in the form of an inequality which is valid for all separable, i.e., non-entangled states, but which may be violated for entangled states. Observing such a violation can then be considered as witnessing the entanglement of the state.

An entanglement witness is typically tailored to specific systems and states. For conventional qbit systems it is common for witnesses to be based on partial tomography, i.e., relying on the determination of individual components of the state. However, like with full state tomography, such methods can be impractical for large numbers of subsystems, and/or subsystems with high dimensionality. It is therefore of interest to derive witnesses which can be detected using ``natural" observables of the system. In the case of heralded entanglement by detection of Raman-scattered photons, as discussed above, it is particularly convenient to use an entanglement witness which can be evaluated by the same type of photodetection measurements as those creating the state. In other words, one may wish to use the detection statistics of subsequent Raman-scattered photons in order to infer entanglement. Since Raman-photon annihilation operators will be in a one-to-one correspondence with the system's creation or annihilation operators, this requires the derivation of inequalities containing system expectation values with an equal number of creation and annihilation operators.

In this article, we derive witnesses particularly designed to detect $W$-state type entanglement in multipartite systems using Raman photodetection statistics. In Section \ref{sec:entwitness}, we first reproduce a known result for the bipartite case ($N=2$) before moving on to an arbitrary number of subsystems $N$. Our main result of this Section is the inequality \eqref{eq:absNMineq} which, if violated, implies that the $N$-partite state in question must involve simultaneous entanglement of more than $M$ parties (in at least one pure component of the state). For $M = N-1$, this allows the detection of genuine $N$-partite entanglement. While the inequality is always violated for the $W$-state described above, this is not necessarily the case if starting from an initial thermal state rather than the ground state. We therefore quantify the witness' robustness by the largest initial thermal occupation for which the witness is violated.

The initial thermal occupation of the subsystems prior to becoming entangled depends on the ratio between the temperature and the excitation frequency of the subsystems. For low-frequency modes, such as in, e.g., engineered mechanical oscillators in optomechanical systems \cite{Aspelmeyer2014RMP}, it can be challenging to satisfy the low initial thermal occupation needed for the general entanglement witness to be useful, even when taking advantage of laser cooling techniques. Motivated by this, we take an alternative approach in Section \ref{sec:nonlocwitness}. Here, we \emph{assume} that the initial detection of a Stokes-photon adds a single excitation to the $N$-partite system. However, we investigate to which degree this excitation is added coherently or incoherently, or in other words, nonlocally or locally. The idea is to test how coherent or nonlocal the measurement back-action of the initial single-photon detection is. The main result of this Section is the inequality \eqref{eq:nonlocbound_distinguishM}, which again can be tested from detection statistics of subsequent Raman-scattered photons. Importantly, we find that this nonlocality witness has a higher tolerance for initial thermal occupation than the general entanglement witness derived in Section \ref{sec:entwitness}.

In Section \ref{sec:measurement}, we show how the witnesses we have derived can be evaluated experimentally. While our results are more generally applicable, we focus in particular on an implementation of this experiment with optomechanical systems. Final discussion and concluding remarks are found in Section \ref{sec:discussion}.

\section{Multipartite Entanglement Witness} \label{sec:entwitness}
Our first result is a general multipartite entanglement witness for systems of independent, identical modes described via ladder operators $b_i$, $b_i^\dagger$. The ladder operators may correspond to a number basis of arbitrary dimension, e.g., Bosonic or Fermionic degrees of freedom. The witness requires neither full nor partial quantum state tomography. It takes the form of a bound on the simple bimodal cross-correlators $\abs{\expval{b_i^\dagger b_j}}$ in terms of number expectation values. The result serves as a generalization of the bipartite entanglement witnesses \cite{HilleryZubairy2006PRL, Borkje2011PRL} to an arbitrary number of independent modes. To derive it, we will exploit that for an arbitrary mixed state, the separability criterion
\begin{equation}\label{eq:sepcriterion}
\expval{O_iO_j}=\expval{O_i}\expval{O_j}
\end{equation}
can be applied to any pure component of that state which is separable in the modes $i$, $j$, where $O_i$, $O_j$ act only on their respective modes.

\subsection{Bipartite Case}
We first consider a system consisting of two independent modes with annihilation operators $b_1$, $b_2$. Suppose the system is in a general separable state
\begin{equation} \label{eq:bipartsepstate}
    \rho = \sum_k P_k \rho_{k1}\otimes \rho_{k2},
\end{equation}
where $\sum_k P_k = 1$ and $\rho_{ki}$ are pure states of the respective modes. Application of the triangle and Cauchy-Schwarz (C-S) inequalities leads to
\begin{subequations}\label{eq:sepbound_convexity}
\begin{equation}
\begin{aligned}
    \abs{\expval{b_1^\dagger b_2}} &\leq \sum_k P_k \abs{ \expval{b_1^\dagger b_2}_k} \\
    &\leq \sum_k P_k \sqrt{\expval{n_1}_k \expval{n_2}_k},
\end{aligned}
\end{equation}
where $n_i = b_i^\dagger b_i$ is the number operator of the $i$'th mode, and $\expval{\cdot}_k = \Tr{\cdot(\rho_{k1}\otimes \rho_{k2})}$. We apply the separability criterion \eqref{eq:sepcriterion} to each term,
\begin{equation}
    \sqrt{\expval{n_1}_k\expval{n_2}_k} = \sqrt{\expval{n_1n_2}_k},
\end{equation}
and find by use of the C-S inequality that the right hand side is concave under mixture,
\begin{equation}
\begin{aligned}
    \sum_k P_k \sqrt{\expval{n_1n_2}_k} &\leq \sqrt{\sum_k P_k} \sqrt{\sum_l P_l \expval{n_1n_2}_l} \\
    &= \sqrt{\sum_l P_l \expval{n_1n_2}_l} = \sqrt{\expval{n_1n_2}}.
\end{aligned}
\end{equation}
\end{subequations}
Thus, any separable state of the form \eqref{eq:bipartsepstate} must satisfy
\begin{equation} \label{eq:abs2ineq}
    \abs{\expval{b_1^\dagger b_2}} \leq \sqrt{\expval{n_1n_2}}.
\end{equation}
Conversely, an entangled state can potentially violate this inequality, allowing it to act as a witness of bipartite entanglement. The inequality \eqref{eq:abs2ineq} is a well known separability criterion by Hillery and Zubairy \cite{HilleryZubairy2006PRL} for witnessing bipartite entanglement of optical modes, which was extended to Raman-scattering oscillator systems in \cite{Borkje2011PRL}.

As an example, consider a Bell-like thermal state 
\begin{equation} \label{eq:Bell-like}
  \rho_c = \frac{1}{2(\nth+1)}(b_1^\dagger+b_2^\dagger)\rho_\text{th.}(b_1+b_2),
\end{equation}
of two harmonic oscillator systems, where $\rho_\text{th.}=\rho_{\text{th.}, 1} \otimes \rho_{\text{th.}, 2}$ is a product state of single mode thermal states
\begin{equation} \label{eq:singlemodethermalstate}
    \rho_{\text{th., i}} = \sum_n \frac{\nth^n}{(\nth+1)^{n+1}} \ket{n}_i \bra{n}_i.
\end{equation}
$\ket{n}_i$ are the Fock states of mode $i$, and $\nth = \Tr{b_i^\dagger b_i \rho_{\text{th., i}}}$ is the average thermal occupation number. The state \eqref{eq:Bell-like} represents the result of a non-local, coherent particle addition process applied to a thermal initial state. Starting from the ground state, i.e., with $\nth=0$, the result $\rho_c = \ket{B}\bra{B}$ with $\ket{B} = \frac{1}{\sqrt{2}}(\ket{1}_1\ket{0}_2+\ket{0}_1\ket{1}_2)$ is a true Bell state. The left hand side of the inequality \eqref{eq:abs2ineq} evaluates to 
\begin{equation}
    \abs{\expval{b_1^\dagger b_2}} = \frac{\nth+1}{2},
\end{equation}
and the right hand side to
\begin{equation}
    \sqrt{\expval{n_1n_2}} = \sqrt{\nth(2\nth+1)}.
\end{equation}
The $\nth$-independent term on the left hand side guarantees that the inequality can always be violated for sufficiently small $\nth$. In this case, violation of \eqref{eq:abs2ineq} occurs for $\nth < \frac{2\sqrt{2}-1}{7} \approx 0.26$.

\subsection{General Case} \label{sec:entwitness_general}
When considering $N$-partite systems, one must distinguish between entanglement involving different numbers of modes. For instance, bipartite entanglement may be detected between every pair of modes in a tripartite system, even if no component of the state has entanglement between all three modes simultaneously. A witness of \emph{genuine} $N$-partite entanglement can instead be found as an inequality which is obeyed by any state with at most $N-1$ simultaneously entangled modes. A reasonable generalization of the separability criterion is therefore to assume that the largest number of simultaneously entangled modes in any term of the mixture is some number $M < N$. This enables the derivation of an $M+1$-partite entanglement witness for $N$-partite systems.

It is useful to first consider the tripartite case $N=3$, $M=2$, which introduces the simplest type of nontrivial separable states, the biseparable states, whose pure components are of the form
\begin{equation} \label{eq:naivebiseparable}
\begin{aligned}
    \rho_{1|2,3} = \rho_{1} \otimes \rho_{2,3},
\end{aligned}
\end{equation}
where one mode is separable from the other two. For the state \eqref{eq:naivebiseparable}, we can apply the bimodal separability criterion \eqref{eq:abs2ineq} to mode pairs $(1, 2)$ and $(1, 3)$, but not to the inseparable pair $(2, 3)$, where one can instead apply a general C-S inequality. This leads to a sum over cross correlators satisfying
\begin{equation} \label{eq:naivebiseparablebound}
\begin{aligned}
    &\abs{\expval{b_1^\dagger b_2}}+\abs{\expval{b_1^\dagger b_3}}+\abs{\expval{b_2^\dagger b_3}} \\
    &\leq \sqrt{\expval{n_1 n_2}}+ \sqrt{\expval{n_1n_3}} + \sqrt{\expval{n_2}\expval{n_3}}.
\end{aligned}
\end{equation}
From the argument \eqref{eq:sepbound_convexity}, this inequality readily applies to mixtures of states of the same form. However, an observed violation indicates only that the state in question must contain terms with a different entanglement structure from \eqref{eq:naivebiseparable}. 

To show that the state contains one or more genuinely tripartite entangled terms, we must produce an inequality which is satisfied for any state
\begin{equation} \label{eq:tripartite_bisep_nonreduced}
    \rho = P_0 \rho_{1|2|3} + P_1 \rho_{1|2,3} + P_2 \rho_{2|1,3} + P_3 \rho_{3|1,2},
\end{equation}
where the first term represents a fully separable state, and the last three represent each of the possible biseparable structures, respectively. Each term may be a mixture of states satisfying the same entanglement structure. However, the fully separable term also satisfies any inequality of the form \eqref{eq:naivebiseparablebound}, and can therefore always be subsumed into the biseparable terms. Thus, we may write the state on the form
\begin{equation} \label{eq:tripartite_bisep_reduced}
    \rho = P_1^\prime \rho^\prime_{1|2,3} + P_2^\prime \rho^\prime_{2|1,3} + P_3^\prime \rho^\prime_{3|1,2},
\end{equation}
where $P_1^\prime+P_2^\prime+P_3^\prime=1$, and each term may contain fully separable components. The primes will be omitted in the following. For this state, we can apply \eqref{eq:naivebiseparablebound} and its permutations component-wise, producing the bound
\begin{equation} \label{eq:tripartiteboundexpansion}
\begin{aligned}
    &\abs{\expval{b_1^\dagger b_2}}+\abs{\expval{b_1^\dagger b_3}}+\abs{\expval{b_2^\dagger b_3}} \\
    &\leq P_1\left(\sqrt{\expval{n_1n_3}_1} + \sqrt{\expval{n_1n_2}_1}\right) \\
    &+P_2\left(\sqrt{\expval{n_2n_3}_2} + \sqrt{\expval{n_1n_2}_2}\right) \\
    &+P_3\left(\sqrt{\expval{n_1n_3}_3} + \sqrt{\expval{n_2n_3}_3}\right) \\
    &+ P_1 \sqrt{\expval{n_2}_1\expval{n_3}_1}+ P_2 \sqrt{\expval{n_1}_2\expval{n_3}_2} + P_3 \sqrt{\expval{n_1}_3\expval{n_2}_3}.
\end{aligned}
\end{equation}
Here we have adopted the subscript convention $\expval{\cdot}_1 = \Tr{\cdot \rho_{1|2,3}}$ etc., where the index refers to the mode which is separable from the rest. Using the inequality of arithmetic and geometric means (AM-GM) gives
\begin{equation}
\begin{aligned}
    \sqrt{\expval{n_1}_l\expval{n_2}_l} &\leq \frac{1}{2}(\expval{n_1}_l + \expval{n_2}_l) \\
    &\leq \frac{1}{2}(\expval{n_1}_l + \expval{n_2}_l + \expval{n_3}_l),
\end{aligned}
\end{equation}
such that the last three terms in \eqref{eq:tripartiteboundexpansion} are bounded by
\begin{equation}
\begin{aligned}
    &P_1 \sqrt{\expval{n_2}_1\expval{n_3}_1}+ P_2 \sqrt{\expval{n_1}_2\expval{n_3}_2} + P_3 \sqrt{\expval{n_1}_3\expval{n_2}_3} \\
    &\leq \frac{1}{2}\sum_{l=1}^3P_l \left( \expval{n_1}_l + \expval{n_2}_l + \expval{n_3}_l\right) \\
    &\equiv \frac{1}{2}\left( \expval{n_1} + \expval{n_2} + \expval{n_3}\right).
\end{aligned}
\end{equation}
The first terms in \eqref{eq:tripartiteboundexpansion} may be treated with the C-S inequality when grouped as follows:
\begin{equation}
\begin{aligned}
    &P_1\sqrt{\expval{n_1n_3}_1} + P_3\sqrt{\expval{n_1n_3}_3}  \\
    &\leq \sqrt{P_1+P_3}\sqrt{P_1\expval{n_1n_3}_1 + P_3\expval{n_1n_3}_3} \\
    &\leq \sqrt{P_1+P_3}\sqrt{\expval{n_1n_3}}.
\end{aligned}
\end{equation}
By using that $\expval{n_1n_3} \leq \max_{i\neq j} \left\{ \expval{n_in_j} \right\}$, the same steps repeated for each pair of modes produces a sum bounded by
\begin{equation} \label{eq:state_reduction_result_tripartite}
\begin{aligned}
    &\left(\sqrt{P_1+P_2} + \sqrt{P_1+P_3}+\sqrt{P_2+P_3}\right) \max_{i\neq j} \left\{ \sqrt{\expval{n_in_j}} \right\} \\
    &\leq 3\sqrt{\frac{2(P_1+P_2+P_3)}{3}} \max_{i\neq j} \left\{ \sqrt{\expval{n_in_j}} \right\} \\
    &=\sqrt{6} \max_{i\neq j} \left\{ \sqrt{\expval{n_in_j}} \right\},
    \end{aligned}
\end{equation}
where we have directly applied concavity of the square root, and equality is obtained for identical probabilities $P_1=P_2=P_3 = \frac{1}{3}$. Note that without undergoing the step from \eqref{eq:tripartite_bisep_nonreduced} to \eqref{eq:tripartite_bisep_reduced}, the additional probability $P_0$ would appear within each square root, resulting in the replacement $\sqrt{6} \rightarrow \sqrt{3P_0+6}$.
%producing the bound $\sqrt{3P_0+6} \leq 3$. 

Combining the above results, the complete witness for genuine tripartite entanglement is
\begin{equation}
\begin{aligned}
    &\sum_{i=1}^2 \sum_{j=i+1}^3 \abs{\expval{b_i^\dagger b_j}} \leq \sqrt{6} \max_{i\neq j} \left\{ \sqrt{\expval{n_in_j}} \right\} + \frac{1}{2}\sum_{i=1}^3 \expval{n_i}.
\end{aligned}
\end{equation}

\begin{figure}
    \includegraphics{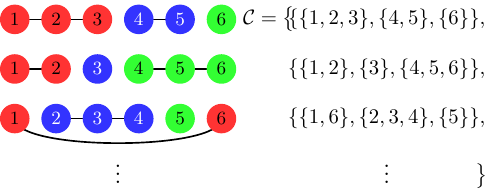}
        \caption{ \label{fig:entanglementstructurehierarchy}
            Every $N$-partite pure state (here with $N=6$) has some structure $S$ of entanglement characterized by disjoint combinations $C$ of modes (illustrated by color and connecting lines) whose modes are inseparable from each other, while modes belonging to different combinations are mutually separable. Structures which are equivalent under exchange of indices, i.e., which contain combinations of the same sizes, belong to the same class $\mathcal{C}$. 
            }
\end{figure}

The generalization of this witness to arbitrary $N$ and $M$ requires the assistance of some more abstract notation. It is useful to introduce the following definitions: 
\begin{itemize}
    \item A \emph{combination} of modes is a subset $C \subseteq \{1, 2, \ldots, N\}$ of the mode indices, representing a particular combination of mutually entangled modes in a given pure state.
    \item A \emph{structure} is a set $S= \{ C_i \}$ of disjoint combinations such that each index is included exactly once, i.e., such that $\bigcup_i C_i = \{1, 2, \ldots, N\}$ and $\bigcap_i C_i = \emptyset$. This represents the overall separability structure of a pure state, with modes belonging to different combinations being separable, and modes belonging to the same combination being inseparable from each other.
    \item A \emph{class} $\mathcal{C}$ of structures denotes a set of structures which is closed under permutation of indices. For a tripartite system, the biseparable structures belong to the same class, and the fully separable structure belongs to its own class.
\end{itemize}
These definitions, representing the subdivision of the modes into internally entangled and mutually separable combinations, allow us to represent arbitrary structures of entanglement of any pure state component. \fref{fig:entanglementstructurehierarchy} provides a visual example of this hierarchy for $N=6$ modes.

% Entanglement structure hierarchy figure
\noindent
\begin{figure}
    \includegraphics{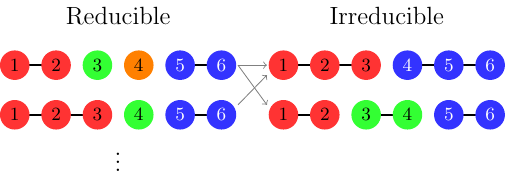}
        \caption{ \label{fig:reduction}
            Given $N=6$ modes and the assumption of maximally $M=3$ simultaneously entangled modes, selective application of the separability criterion \eqref{eq:abs2ineq} allows the structures on the left to be treated as if they were one of the structures on the right, reducing the number of separable cross correlator pairs. Of the reducible structures, the topmost may be treated as either of the irreducible structures, while the second one may only be treated as the topmost irreducible structure. For the irreducible structures, no further reduction is appropriate, as combining any two combinations leads to entanglement of more than $M=3$ modes. Every choice of $N$ and $M$ leads to some collection of irreducible classes of structures. $N=6$, $M=3$ produces the two above.
            }
\end{figure}

Naturally, any state can be represented as a mixture whose terms are arranged by structure and class:
\begin{equation}
\begin{aligned}
    \rho &= \sum_\mathcal{C} \sum_{S \in \mathcal{C}} P_S \rho_S.
\end{aligned}
\end{equation}
Using this decomposition, we can bound the sum over cross correlator pairs by
\begin{equation} \label{eq:NMineq_basic}
\begin{aligned}
    \sum_\text{pairs} \abs{\expval{b_i^\dagger b_j}} &\leq \sum_\text{pairs} \sum_\mathcal{C} \sum_{S \in \mathcal{C}: i|j} P_S \sqrt{\expval{n_in_j}_S} \\
    &+ \sum_\text{pairs} \sum_\mathcal{C} \sum_{S \in \mathcal{C}: i\sim j} P_S \sqrt{\expval{n_i}_S\expval{n_j}_S},
\end{aligned}
\end{equation}
where $i|j$ signifies that the modes $i$ and $j$ are separable in $S$, and $i\sim j$ represents inseparability. As with the tripartite derivation, where we could treat the fully separable terms as if they were biseparable, we may again treat terms belonging to some classes as if they belong to others.
In general, if a structure contains two combinations $C_a, C_b \in S$ such that $\abs{C_a}+\abs{C_b} \leq M$, it is subject to the same inequality as the structure where the combinations are replaced with their union $C_a, C_b \to C_a \cup C_b$. ``Reducible" terms, for which this is the case, can thereby be absorbed into terms where any such action would violate the assumption of maximally $M$ simultaneous entangled modes, leaving a collection of ``irreducible" terms. This process ensures the number of separable pairs which arise in the cross correlator sum is minimized, without affecting the treatment of the inseparable pairs. An example of reducible entanglement classes and corresponding irreducible classes is shown in \fref{fig:reduction} for $N=6$ and $M=3$.

% Reduction of state
\noindent
After representing the state in terms of the irreducible classes, we may apply the C-S inequality to the remaining separable terms, producing
\begin{equation}
\begin{aligned}
    &\sum_\text{pairs} \sum_\mathcal{C} \sum_{S \in \mathcal{C}: i|j} P_S \sqrt{\expval{n_in_j}_S} \\
    &\leq \sum_\text{pairs} \sqrt{\sum_\mathcal{C} \sum_{S \in \mathcal{C}: i|j} P_S \expval{n_in_j}_S} \sqrt{\sum_\mathcal{C} \sum_{S \in \mathcal{C}: i|j} P_S} \\
    &\leq \sum_\text{pairs} \sqrt{\expval{n_in_j}} \sqrt{\sum_\mathcal{C} \sum_{S \in \mathcal{C}: i|j} P_S} \\
    &\leq \max_{i\neq j} \left\{ \sqrt{\expval{n_in_j}} \right\} \sum_\text{pairs} \sqrt{\sum_\mathcal{C} \sum_{S \in \mathcal{C}: i|j} P_S}.
\end{aligned}
\end{equation}
Concavity of the square root can then be applied to the remaining sum, producing the maximum
\begin{equation} \label{eq:sqrtconcavity_NMsep}
\begin{aligned}
    \sum_\text{pairs} \sqrt{\sum_\mathcal{C} \sum_{S \in \mathcal{C}: i|j} P_S} &\leq N_\text{pairs} \sqrt{ \frac{1}{N_\text{pairs}}\sum_\text{pairs }\sum_\mathcal{C} \sum_{S \in \mathcal{C}: i|j} P_S},
\end{aligned}
\end{equation}
where $N_\text{pairs}\equiv \binom{N}{2} = N(N-1)/2$ is the number of unique pairs selectable from $N$ modes. Next we use that
\begin{equation} \label{eq:NsepC}
\begin{aligned}
    \sum_\text{pairs}\sum_{S \in \mathcal{C}: i|j} P_S &= \sum_{S \in \mathcal{C}} P_S \sum_{\text{pairs}:i|j\text{ for }S} 1 \\
    &\equiv \sum_{S \in \mathcal{C}} P_S N_{\text{sep.},\mathcal{C}}.
\end{aligned}
\end{equation}
$N_{\text{sep.},\mathcal{C}}$ is the total number of unique mode pairs which are separable for a structure in the class $\mathcal{C}$. Defining the probability $P_\mathcal{C} \equiv \sum_{S \in \mathcal{C}} P_S$ of the state having a structure belonging to class $\mathcal{C}$, we write the right hand side of \eqref{eq:sqrtconcavity_NMsep} as
\begin{equation}
\begin{aligned}
    &\sqrt{ N_\text{pairs} \sum_\mathcal{C} P_\mathcal{C} N_{\text{sep.},\mathcal{C}}}.
\end{aligned}
\end{equation}
Ambiguity with respect to what the probabilities $P_\mathcal{C}$ are necessitates another step of maximization. The maximum is obtained when $P_\mathcal{C} = 1$ for the class which results in the highest value of $N_{\text{sep.},\mathcal{C}}$:
\begin{equation} \label{eq:NsepCmax}
\begin{aligned}
    \sum_\mathcal{C} P_\mathcal{C} N_{\text{sep.},\mathcal{C}} \leq \max_\mathcal{C} \{ N_{\text{sep.},\mathcal{C}} \} \equiv N_{\text{sep.},\mathcal{C}, \text{max}}.
\end{aligned}
\end{equation}
Thus we have
\begin{equation} \label{eq:sep_terms_ineq}
\begin{aligned}
    &\sum_\text{pairs} \sum_\mathcal{C} \sum_{S \in \mathcal{C}: i|j} P_S \sqrt{\expval{n_in_j}_S} \\
    &\leq \max_{i\neq j} \left\{ \sqrt{\expval{n_in_j}} \right\} \sqrt{N_\text{pairs} N_{\text{sep.},\mathcal{C}, \text{max}}}.
\end{aligned}
\end{equation}
A general expression for $N_{\text{sep.},\mathcal{C}, \text{max}}$ is derived in App. \ref{app:Finding_NsepCmax}, and values for various choices of $N$ and $M$ are shown in \tref{tab:N_sepCmax}. We note that, in general, this quantity has a rather complicated dependence on the given choice of $N$ and $M$. However, for the choice $M=N-1$, corresponding to genuine multipartite entanglement, all ``irreducible" structures contain only two combinations, and one can therefore easily compute
\begin{equation}
    N_{\text{sep.},\mathcal{C}, \text{max}}[M=N-1] = \begin{cases}
        \frac{N^2}{4}, & \ N\text{ even}, \\
        \frac{N^2-1}{4}, & \ N\text{ odd},
    \end{cases} 
\end{equation}
corresponding to combinations whose size are equal ($N$ even) or differ by one ($N$ odd).

The sum in \eqref{eq:NMineq_basic} which runs over inseparable pairs can be treated more straightforwardly. Similarly to the tripartite case, we apply the AM-GM inequality and express the sum as
\begin{equation}
\begin{aligned}
    &\sum_\text{pairs} \sum_\mathcal{C} \sum_{S \in \mathcal{C}: i\sim j} P_S \sqrt{\expval{n_i}_S\expval{n_j}_S} \\
    &\leq \frac{1}{2} \sum_\text{pairs} \sum_\mathcal{C} \sum_{S \in \mathcal{C}: i\sim j} P_S (\expval{n_i}_S + \expval{n_j}_S) \\
    &= \frac{1}{2} \sum_\mathcal{C} \sum_{S \in \mathcal{C}} P_S \sum_{\text{pairs}:i\sim j\text{ for }S} (\expval{n_i}_S + \expval{n_j}_S).
\end{aligned}
\end{equation}
In the innermost sum, each expectation value $\expval{n_i}_S$ occurs exactly $\abs{C(S, i)}-1$ times, where $C(S, i)$ is the combination of modes in structure $S$ which contains that index, and $\abs{\cdot}$ denotes the number of modes in that combination. To symmetrize this sum such that an expectation value of the complete state is recovered, we bound $\abs{C(S, i)}$ by $M$ for every combination, and obtain
\begin{equation}
\begin{aligned}
    &\frac{1}{2} \sum_\mathcal{C} \sum_{S \in \mathcal{C}} P_S \sum_{\text{pairs}:i\sim j\text{ for }S} (\expval{n_i}_S + \expval{n_j}_S) \\
    &\leq \frac{1}{2} \sum_\mathcal{C} \sum_{S \in \mathcal{C}} P_S \sum_{i=1}^N(M-1)\expval{n_i}_S\\
    &= \frac{M-1}{2}\sum_{i} \sum_\mathcal{C} \sum_{S \in \mathcal{C}} P_S \expval{n_i}_S = \frac{M-1}{2}\sum_{i=1}^N \expval{n_i}.
\end{aligned}
\end{equation}
Thus, the inseparable terms always satisfy
\begin{equation} \label{eq:insep_terms_ineq}
\begin{aligned}
    \sum_\text{pairs} \sum_\mathcal{C} \sum_{S \in \mathcal{C}: i\sim j} P_S \sqrt{\expval{n_i}_S\expval{n_j}_S} \leq \frac{M-1}{2} \sum_{i=1}^N \expval{n_i}.
\end{aligned}
\end{equation}
\eqref{eq:NMineq_basic} combined with \eqref{eq:sep_terms_ineq} and \eqref{eq:insep_terms_ineq} then gives the final entanglement witness
\begin{equation} \label{eq:absNMineq}
\begin{aligned}
    &\sum_\text{pairs} \abs{\expval{b_i^\dagger b_j}} \\
    &\leq \max_{i\neq j} \left\{ \sqrt{\expval{n_in_j}} \right\} \sqrt{N_\text{pairs} N_{\text{sep.},\mathcal{C}, \text{max}}} + \frac{M-1}{2} \sum_{i=1}^N \expval{n_i}.
\end{aligned}
\end{equation}
Violation implies that the state must involve simultaneous entanglement of more than $M$ modes, i.e., genuine $M+1$-partite entanglement.

\subsection{Example: Thermal \texorpdfstring{$W$}{W}-like state}
We can now apply the witness \eqref{eq:absNMineq} to a $W$-like excited thermal state
\begin{equation} \label{eq:NpartiteWlikestate}
  \rho_c = \frac{1}{N(\nth+1)}\sum_{ij} b_i^\dagger \rho_\text{th.} b_j
\end{equation}
of $N$ harmonic oscillators, where
\begin{equation} \label{eq:ThermalState}
    \rho_\text{th.} = \bigotimes_{i=1}^N \rho_{\text{th.}, i}
\end{equation}
is a multimode thermal state with $\rho_{\text{th.}, i}$ as defined in \eqref{eq:singlemodethermalstate}. The state \eqref{eq:NpartiteWlikestate} corresponds to coherent/nonlocal particle addition to an initial thermal state across all $N$ harmonic oscillators. In the case of an initial ground state, $\nth = 0$, this results in the true $W$-state $\rho_c = \ket{W_N}\bra{W_N}$ with $\ket{W_N} = \frac{1}{\sqrt{N}}(\ket{1,0,\ldots}+\ket{0,1,\ldots}+\ldots+\ket{0,\ldots,1})$.
The state \eqref{eq:NpartiteWlikestate} produces the expectation values
\begin{align} \label{eq:Wlike_expvals}
    \abs{\expval{b_i^\dagger b_j}_c} &= \frac{\nth + 1}{N}, \\
    \expval{n_i}_c &= \frac{1}{N}\left( N\nth + (\nth + 1)\right) \\
    \expval{n_i n_j}_c &= \frac{\nth}{N} (N \nth + 2(\nth+1))
\end{align}
for all $i \neq j$ due to the symmetry of the state. There are $N_\text{pairs} = N(N-1)/2$ terms on the left hand side of \eqref{eq:absNMineq}, and the inequality becomes
\begin{equation} \label{eq:threshold_sym_thermal}
\begin{aligned}
     &\frac{N-1}{2} (\nth + 1) \\
     &\leq \sqrt{\frac{(N-1)}{2} \nth (N \nth + 2(\nth+1)) N_{\text{sep.},\mathcal{C}, \text{max}}}\\
     &+\frac{M-1}{2}\left( N\nth + (\nth + 1)\right),
\end{aligned}
\end{equation}
where the maximal possible number of separable pairs $N_{\text{sep.},\mathcal{C}, \text{max}}$ for any given component was defined in \eqref{eq:NsepCmax} (see also App. \ref{app:Finding_NsepCmax}) and whose values for $N \leq 7$ can be found in \tref{tab:N_sepCmax}. For thermal occupation $\nth=0$, the inequality reduces to
%\begin{equation}
%    \frac{N-1}{2} \leq \frac{M-1}{2},
%\end{equation}
\begin{equation}
    N \leq M,
\end{equation}
%which is always violated for any choice $N\geq 2$, $0 < M < N$. 
which is always violated for any choice $M < N$. It is therefore possible to witness entanglement of a $W$-like state to every ``order" $M+1$ given sufficiently low thermal occupation. \tref{tab:sym_thresholds} shows the thermal thresholds below which violation occurs as a function of $N$ and $M$.

\begin{figure}[tb]
	% \begin{tikzpicture}[scale=1]
	% \begin{semilogyaxis}[xlabel=$N$, ylabel={$\nth$}, ymax=0.33, xticklabel style={anchor=north}, xmin=1.99, xmax = 10, xtick={2,3,4,5,6,7,8,9,10}, grid=both]
 %    \addplot+[color=blue] table [x index=0,y index=1,col sep=comma] {Plots/thresh_abswitness_trick_ferm.csv};
	% \addplot+[color=red] table [x index=0,y index=1,col sep=comma] {Plots/thresh_abswitness_trick.csv};
	% \end{semilogyaxis}
	% \end{tikzpicture}
    \includegraphics{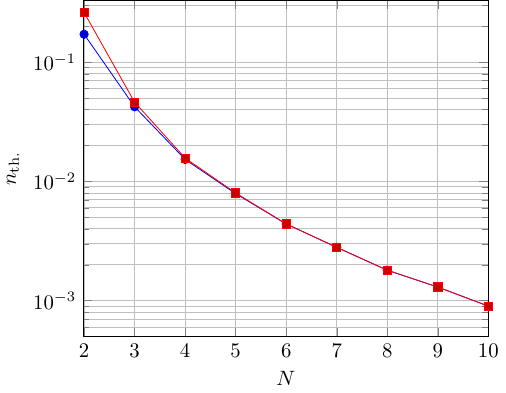}
	\caption{\label{fig:thresholds_abswitness_true}
	    Thermal occupation numbers $\nth$ which saturate the inequality \eqref{eq:threshold_sym_thermal} (in red) evaluated for $M=N-1$, corresponding to genuine $N$-partite entanglement. For thermal occupations below these thresholds, violation of the entanglement witness \eqref{eq:absNMineq} occurs, given a thermal $W$-like state. The corresponding values for an ensemble of two-level systems is shown in blue. It is clear that for increasing $N$, the values converge. This is because the $W$-like states for harmonic oscillators and two-level systems give identical number statistics in the limit $\nth \to 0$, where no components of two or more excitations are present for a harmonic oscillator.
		}
\end{figure}

Alternatively, our witness can be adapted to two-level systems without loss of generality, by replacing the Bosonic ladder operators with corresponding Fermionic ladder operators. Specifically, the single mode thermal state now has the form
\begin{equation}
    \rho_{\text{th.}, \text{qbit}} = (1-\nth)\ket{0}\bra{0}+\nth\ket{1}\bra{1},
\end{equation}
with $\nth \leq 1$, where particle addition now annihilates the $1$-particle component (representing that this component does not contribute to the particle addition process). The expectation values become
\begin{align} \label{eq:Wlike_expvals_2level}
    \abs{\expval{b_i^\dagger b_j}_{c,\text{qbit}}} &= \frac{1-\nth}{N}, \\
    \expval{n_i}_{c,\text{qbit}} &= \frac{1}{N}\left( N\nth + (1-\nth)\right) \\
    \expval{n_i n_j}_{c,\text{qbit}} &= \frac{\nth}{N} (N \nth + 2(1-\nth)),
\end{align}
and the two-level version of \eqref{eq:threshold_sym_thermal} is
\begin{equation} \label{eq:threshold_sym_thermal_2level}
\begin{aligned}
     &\frac{N-1}{2} (1-\nth) \\
     &\leq \sqrt{\frac{(N-1)}{2} \nth (N \nth + 2(1-\nth)) N_{\text{sep.},\mathcal{C}, \text{max}}}\\
     &+\frac{M-1}{2}\left( N\nth + (1-\nth)\right).
\end{aligned}
\end{equation}
The thermal thresholds for violation of the witnesses for genuine $N$-partite entanglement ($M=N-1$) are shown in \fref{fig:thresholds_abswitness_true}.

\begin{figure}[tb]
% 	\begin{tikzpicture}[scale=1]
% 	\begin{semilogxaxis}[xlabel=$\nth$, ylabel={$M+1$}, ymin = 1.7, ymax = 17, xticklabel style={anchor=north}, xmin=3e-3, xmax = 0.5, ytick={0,1,...,30}, grid=both, legend image post style={scale=0.5}, legend style={font=\footnotesize}]
%     \legend{TLS: $N=N_\text{opt.}(M)$, HO: $N=N_\text{opt.}(M)$, TLS: $N=M+1$, HO: $N=M+1$};
%     \addplot[color=blue, thick] table [x index=0,y index=1,col sep=comma] {Plots/absNoptfermvals.csv};
%     \addplot[color=red, thick] table [x index=0,y index=1,col sep=comma] {Plots/absNoptvals.csv};
%     \addplot[color=blue, dashed, thick] table [x index=0,y index=1,col sep=comma] {Plots/absNfermvals.csv};
%     \addplot[color=red, dashed, thick] table [x index=0,y index=1,col sep=comma] {Plots/absNvals.csv};

%     \addplot[draw=none, point meta =explicit,
% nodes near coords, every node near coord/.style={black!50, yshift=0pt, xshift=5pt}] table [x index=1,y index=0, meta index =2, col sep=comma] {Plots/thresh_abswitness_trick_optN.csv};
% 	\end{semilogxaxis}
% 	\end{tikzpicture}
    \includegraphics{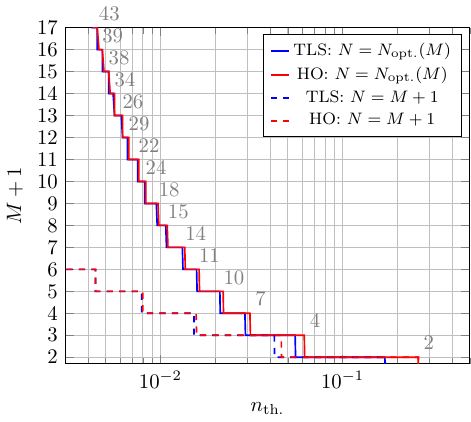}
	\caption{\label{fig:achievableMplus1_Nopt_entwitness}
	    The highest order of entanglement $M+1$ for which the inequalities (\eqref{eq:threshold_sym_thermal} in solid red, \eqref{eq:threshold_sym_thermal_2level} in solid blue) are violated for a thermal $W$-like state at a given thermal occupation $\nth$. The optimal number of subsystems $N_\text{opt.}$ is indicated by numbers above their corresponding horizontal grid lines, and are identical between the harmonic oscillator and two-level system cases. The values for the case $N=M+1$ are shown in dashed lines for comparison.
		}
\end{figure}

From \tref{tab:sym_thresholds}, one can notice that the maximal thermal threshold for a given order of entanglement $M+1$ occurs for some optimal number of subsystems $N=N_\text{opt.} \geq M+1$. One can therefore consider the case where the thermal occupation $\nth$ is fixed and $N$ can be varied to maximize the order $M+1$ of entanglement which can be witnessed at that temperature. \fref{fig:achievableMplus1_Nopt_entwitness} shows the achievable degrees of entanglement as a function of $\nth$ provided that the optimal number of subsystems can be realized. In App. \ref{app:Opt_N}, $N_\text{opt.}$ is found as a function of $M$, and shown to be roughly linear.

\begin{table}
    \centering
    \begin{tabular}{|p{20pt}|p{20pt}|p{20pt}|p{20pt}|p{20pt}|p{20pt}|p{20pt}|}
        \hline  &  M=1 & 2 & 3 & 4 & 5 & 6 \\\hline
        N=2 & 1 &  &  & & & \\ \hline
        3 & 3 & 2 & & & & \\ \hline
        4 & 6 & 4 & 4 & & & \\ \hline 
        5 & 10 & 8 & 6 & 6 &  & \\ \hline
        6 & 15 & 12 & 12 & 9 & 9 & \\ \hline
        7 & 21 & 18 & 16 & 12 & 12 & 12 \\ \hline
    \end{tabular}
    \caption{The maximal number of separable pairs $N_{\text{sep.},\mathcal{C},\text{max}}$ as defined in \eqref{eq:NsepC} and \eqref{eq:NsepCmax}, as a function of the number of modes $N$ and the assumed maximal number of simultaneously entangled modes $M$.}
    \label{tab:N_sepCmax}
\end{table}

\begin{table}
    \centering
    \begin{tabular}{|p{20pt}|p{20pt}|p{20pt}|p{20pt}|p{20pt}|p{20pt}|p{20pt}|}
        \hline  &  M=1 & 2 & 3 & 4 & 5 & 6 \\\hline
        N=2 & 0.261 &  &  & & & \\ \hline
        3 & 0.160 & 0.046 & & & & \\ \hline
        4 & 0.116 & 0.062 & 0.016 & & & \\ \hline 
        5 & 0.090 & 0.055 & 0.029 & 0.008 &  & \\ \hline
        6 & 0.074 & 0.052 & 0.028 & 0.016 & 0.004 & \\ \hline
        7 & 0.063 & 0.046 & 0.031 & 0.022 & 0.010 & 0.003 \\ \hline
    \end{tabular}
    \caption{Thermal thresholds producing equality in \eqref{eq:threshold_sym_thermal} for violation of the entanglement witness \eqref{eq:absNMineq} given $N$ harmonic oscillators in a thermal $W$-like state, as a function of the number of modes $N$ and the assumed maximal number of simultaneously entangled modes $M$.}
    \label{tab:sym_thresholds}
\end{table}

\section{Witness of Nonlocal Backaction} \label{sec:nonlocwitness}
In the previous sections we have treated entanglement as a property of a given multipartite state; witnessing entanglement means to interrogate the state and receive an answer which a separable state would be incapable of providing. An alternative approach is to investigate to what extent specific processes or events can generate entanglement. For example, a key question to address is whether the back-action due to a single-photon detection is distributed quantum coherently across the subsystems, rather than incoherently/probabilistically as in a classical system. In the case of $W$-state preparation this means to verify that the projective single excitation process, which ideally should look like
\begin{equation} \label{eq:1PAgeneral}
    \rho_0 \to \rho_c = \frac{b_W^\dagger \rho_0 b_W}{\Tr{b_W^\dagger \rho_0 b_W}},
\end{equation}
is nonlocal, where
\begin{equation}
  b_W = \sum_i c_i b_i
\end{equation}
is an arbitrary collective oscillator mode, and $\rho_0$ is an arbitrary initial state. A witness of the nonlocality of this process can be derived on similar principles as the conventional witness in Sec. \ref{sec:entwitness}. In place of the generalized separability criterion, we introduce the assumption of semi-locality, i.e., that the process acts on a subset of the modes rather than all $N$ modes as expected. Physically this can occur if, e.g., some part of the system or environment acts like a detector which can distinguish between which modes (or groups of modes) the excitation may have been added to, in general leading to a mixture of projections each involving ladder operators of fewer than $N$ modes.

\subsection{Bipartite Case}
For $N=2$ modes, the process \eqref{eq:1PAgeneral} coherently adds one particle to the system which is localized to neither mode $1$ or mode $2$. Given that a single particle was indeed added, but nonlocality of the form \eqref{eq:1PAgeneral} was not achieved, the only alternative is if the particle was added locally to \emph{either} one mode \emph{or} the other, leading to the mixture
\begin{equation} \label{eq:bipartite_loc}
    \rho_0 \to \rho_c = P_1 \rho_{c,1} + P_2 \rho_{c,2},
\end{equation}
where
\begin{equation} \label{eq:bipartite_loc_comps}
    \rho_{c,i} = \frac{b_i^\dagger \rho_0 b_i}{\Tr{b_i^\dagger \rho_0 b_i}},
\end{equation}
and
\begin{equation} \label{eq:bipartite_loc_probs}
    P_i = \frac{\abs{c_i}^2 \Tr{b_i^\dagger \rho_0 b_i}}{\abs{c_1}^2 \Tr{b_1^\dagger \rho_0 b_1}+\abs{c_2}^2 \Tr{b_2^\dagger \rho_0 b_2}}
\end{equation}
is the probability that the particle was added to mode $i$, and satisfies $P_1 + P_2 = 1$. We will assume that the amplitudes $\abs{c_i}^2$ can be independently measured. For example, in experiments detecting single Raman-scattered photons (see Sec. \ref{sec:measurement}), this would involve the measurement of transmission amplitudes of a linear optical network. 

Before we begin, we note that the following derivation depends on the structure of the underlying Hilbert space, and we assume throughout that the system obeys Bosonic commutation relations. An alternative derivation for Fermionic (i.e., two-level) systems is summarized in Sec. \ref{sec:nonlocwitness_therm_example}.
%, but not produced in detail. 
As with our conventional entanglement witness, the strategy is to find a limit to the simple cross correlations which the state \eqref{eq:bipartite_loc} cannot exceed, but which may be violated by the state \eqref{eq:1PAgeneral}. We consider the decomposition of the cross correlator,
\begin{equation} \label{eq:loc_bipart_decomp}
    \expval{b_1^\dagger b_2}_c = P_1 \expval{b_1^\dagger b_2}_{c1}+P_2 \expval{b_1^\dagger b_2}_{c2},
\end{equation}
where $\expval{\cdot}_c = \Tr{\cdot \rho_c}$ and $\expval{\cdot}_{ci} = \Tr{\cdot \rho_{c,i}}$. For each of the above terms, we can rewrite the expectation values as, e.g.,
\begin{equation} \label{eq:rel_trick}
\begin{aligned}
    &\expval{b_1^\dagger b_2}_{c1} = \frac{\expval{\left(b_1 b_1^\dagger - \expval{b_1 b_1^\dagger}_0 \right) b_2 b_1^\dagger}_0}{\expval{b_1 b_1^\dagger}_0} +  \expval{b_1^\dagger b_2}_0,
\end{aligned}
\end{equation}
where $\expval{\cdot}_0 = \Tr{\cdot \rho_0}$ denotes expectation values of the initial state. Application of the C-S inequality then produces the bound
\begin{equation}
    \abs{\expval{b_1^\dagger b_2}_{c1}-\expval{b_1^\dagger b_2}_0} \leq \frac{\sqrt{\expval{\Delta n_1^2}_0 (\expval{n_1 n_2}_0 + \expval{n_2}_0)}}{\expval{n_1}_0+1}.
\end{equation}
Combining this with an equivalent treatment of the second term in \eqref{eq:loc_bipart_decomp} gives
\begin{equation}
\begin{aligned}
    \abs{\expval{b_1^\dagger b_2}_{c} -\expval{b_1^\dagger b_2}_0} &\leq \frac{P_1 \sqrt{\expval{\Delta n_1^2}_0 (\expval{n_1 n_2}_0 + \expval{n_2}_0)}}{\expval{n_1}_0+1} \\
    &+ \frac{P_2 \sqrt{\expval{\Delta n_2^2}_0 (\expval{n_1 n_2}_0 + \expval{n_1}_0)}}{\expval{n_2}_0+1}.
\end{aligned}
\end{equation}
With the probabilities \eqref{eq:bipartite_loc_probs} substituted in, we get the final bound
\begin{equation} \label{eq:bipart_nonloc_bound}
\begin{aligned}
    \abs{\expval{b_1^\dagger b_2}_{c} -\expval{b_1^\dagger b_2}_0} &\leq \frac{1}{\abs{c_1}^2 (\expval{n_1}_0+1) + \abs{c_2}^2 (\expval{n_2}_0 + 1)}\\
    &\times \bigg( \abs{c_1}^2 \sqrt{\expval{\Delta n_1^2}_0 (\expval{n_1 n_2}_0 + \expval{n_2}_0)} \\
    &+ \abs{c_2}^2 \sqrt{\expval{\Delta n_2^2}_0 (\expval{n_1 n_2}_0 + \expval{n_1}_0)} \bigg).
\end{aligned}
\end{equation}
Note that, unlike the entanglement witness \eqref{eq:absNMineq} where all expectation values refer to the state in question, here the expectation value $\expval{b_1^\dagger b_2}_c$ of the conditioned state $\rho_c$ is compared to expectation values in the initial state $\rho_0$.

For two harmonic oscillators in a thermal initial state $\rho_0 = \rho_\text{th.} \otimes \rho_\text{th.}$, using that $\expval{n_i}_\text{th.} = \nth$, $\expval{n_i n_j}_\text{th.} = \nth^2$, and $\expval{\Delta n_i^2}_\text{th.} = \nth(\nth+1)$, the bound \eqref{eq:bipart_nonloc_bound} reduces to
\begin{equation} \label{eq:bipart_nonloc_therm}
    \nth \geq 1,
\end{equation}
which is obviously violated when the initial thermal occupation satisfies $\nth < 1$. 

Violation of the inequality \eqref{eq:bipart_nonloc_bound} implies that, given the assumption that a single particle is added to the state, it must happen coherently, or nonlocally. In order to claim that $\rho_c$ is entangled, one must also assume a separable initial state $\rho_0$ (or that, if $\rho_0$ is entangled, it belongs to a set of states for which entanglement cannot be removed by single-particle addition). However, for remote systems (see Sec. \ref{sec:measurement}), the assumption of separable $\rho_0$ can be a very natural one.

\subsection{General Case}
For an $N$-partite system with $N> 2$, a particle may in principle be coherently added to any given combination of modes of size less than or equal to $N$. Ambiguity about which such combination the particle was added to results in mixtures of particle-added states, whose components can be nonlocal to mutually nonintersecting combinations of modes. The formalism introduced in Sec. \ref{sec:entwitness_general} is therefore well suited to treating $N$-partite nonlocality. Letting, as before, a structure $S$ be a set of nonintersecting combinations $C$ of modes such that $C_1 \cup C_2 \cup \ldots = \{1,2,\ldots,N\}$, the process acts nonlocally onto any one of the combinations in the structure, with some probability $P_C$. The structure may contain a single combination of size $N$, corresponding to full nonlocality, or it may contain $N$ combinations of size $1$, corresponding to full locality. If the largest combination of the structure is of size $M$, where $1 < M < N$, we say the process acts semilocally. We may therefore characterize the degree of nonlocality of the process by this maximal combination size. 

For a particular structure, the semilocal particle addition model takes the form
\begin{equation} \label{eq:1PAsemilocal}
    \rho_0 \to \rho_{c} = \sum_{C \in S} P_C \frac{b_C^\dagger \rho_0 b_C}{\Tr{b_C^\dagger \rho_0 b_C}},
\end{equation}
where $b_C = \sum_{i \in C} c_i b_i/\sqrt{\sum_{i \in C} \abs{c_i}^2}$ is a normalized truncation of the mode $b_W$ (in the fully nonlocal model \eqref{eq:1PAgeneral}) to the subset of modes in $C$, and
\begin{equation} \label{eq:general_loc_probs}
    P_C = \frac{ \Tr{b_C^\dagger \rho_0 b_C} \sum_{i \in C} \abs{c_i}^2}{\sum_{C^\prime \in S} \Tr{b_{C^{\prime}}^\dagger \rho_0 b_{C^{\prime}}} \sum_{i \in C^\prime} \abs{c_i}^2}
\end{equation}
is the probability that the particle was added to combination $C$, with $\sum_{C \in S} P_C = 1$.

We consider a conditioned simple cross-correlator of the state \eqref{eq:1PAsemilocal},
\begin{equation}
    \expval{b_1^\dagger b_2}_c = \sum_{C \in S}  P_C \frac{\expval{b_C b_1^\dagger b_2 b_C^\dagger}_0}{\expval{b_C b_C^\dagger}_0}.
\end{equation}
The trick \eqref{eq:rel_trick} can be applied when $b_C b_C^\dagger$ can be collected on either side of the expectation value in the numerator. For combinations $C$ which do not include both modes $1$ and $2$, we can commute either $b_C$ to the right or $b_C^\dagger$ to the left, leading to the bounds
\begin{subequations} \label{eq:nonloc_mainterms}
\begin{equation}
    \abs{\frac{\expval{b_C b_1^\dagger b_2 b_C^\dagger}_0}{\expval{b_C b_C^\dagger}_0}-\expval{b_1^\dagger b_2}_0} \leq \frac{\sqrt{\expval{\Delta n_C^2}_0 (\expval{n_1 n_2}_0 + \expval{n_2}_0)}}{\expval{n_C}_0+1}
\end{equation}
for commutation to the left ($C$ does not contain mode $2$), and 
\begin{equation}
    \abs{\frac{\expval{b_C b_1^\dagger b_2 b_C^\dagger}_0}{\expval{b_C b_C^\dagger}_0}-\expval{b_1^\dagger b_2}_0} \leq \frac{\sqrt{\expval{\Delta n_C^2}_0 (\expval{n_1 n_2}_0 + \expval{n_1}_0)}}{\expval{n_C}_0+1}
\end{equation}
\end{subequations}
for commutation to the right ($C$ does not contain mode $1$). $n_C \equiv b_C^\dagger b_C$ is the number operator for the collective mode, and $\Delta n_C^2$ is its variance. For cases where neither mode is in $C$ we can choose commutation direction arbitrarily, and we will choose to average the two alternatives. 

The remaining choices of $C$ contain both modes $1$ and $2$. For these terms, commutation to either side produces non-zero commutators
\begin{align}
    \comm{b_1^\dagger b_2}{b_C^\dagger} &= \frac{c_{2}^*}{\sqrt{\sum_{i \in C} \abs{c_i}^2}} b_1^\dagger \equiv c_{C,2}^* b_1^\dagger, \\
    \comm{b_C}{b_1^\dagger b_2} &= \frac{c_{1}}{\sqrt{\sum_{i \in C} \abs{c_i}^2}} b_2 \equiv c_{C,1} b_2.
\end{align}
We choose to average between commutation directions here as well, and, by the triangle inequality, we obtain
\begin{equation} \label{eq:nonloc_with_comm}
\begin{aligned}
    &\abs{\frac{\expval{b_C b_1^\dagger b_2 b_C^\dagger}_0}{\expval{b_C b_C^\dagger}_0}-\expval{b_1^\dagger b_2}_0} \\
    &\leq \frac{\sqrt{\expval{\Delta n_C^2}_0}}{2(\expval{n_C}_0+1)} \left(\sqrt{\expval{n_1 n_2}_0 + \expval{n_2}_0} + \sqrt{\expval{n_1 n_2}_0 + \expval{n_1}_0} \right)\\
    &+ \frac{1}{2(\expval{n_C}_0+1)}\left(\abs{c_{C,2}} \abs{ \expval{b_C b_1^\dagger}_0} +\abs{c_{C,1}} \abs{ \expval{b_2 b_C^\dagger}_0}\right).
\end{aligned}
\end{equation}
By commuting the operators in the last line, and subsequently applying the triangle and C-S inequalities, we obtain 
\begin{equation} 
\begin{aligned}
    &\abs{c_{C,2}} \abs{ \expval{b_C b_1^\dagger}_0} +\abs{c_{C,1}} \abs{ \expval{b_2 b_C^\dagger}_0} \\
    &\leq \abs{c_{C,2}} \abs{ \expval{ b_1^\dagger b_C}_0} +\abs{c_{C,1}} \abs{ \expval{ b_C^\dagger b_2}_0} + 2 \abs{c_{C,1}}\abs{c_{C,2}} \\
    &\leq \abs{c_{C,2}} \sqrt{ \expval{ n_1}_0 \expval{ n_C}_0} +\abs{c_{C,1}} \sqrt{ \expval{ n_2}_0 \expval{ n_C}_0} + 2 \abs{c_{C,1}}\abs{c_{C,2}}.
\end{aligned}
\end{equation}

So far, we have only considered $\expval{b_1^\dagger b_2}$. For the sum $\sum_\text{pairs}\expval{b_i^\dagger b_j}$, terms like in the last line of \eqref{eq:nonloc_with_comm} combine to the sum
\begin{equation}
    \frac{1}{2(\expval{n_C}_0+1)}\sum_{\substack{i,j \in C: \\ j\neq i}} \abs{c_{C,j}}\left(\sqrt{ \expval{ n_i}_0 \expval{ n_C}_0} + \abs{c_{C,i}}\right).
\end{equation}
To eliminate the coefficients $c_{C,i}$, we can apply the C-S inequality to the sums over $j$, using
\begin{equation}
\begin{aligned}
    \sum_{\substack{j \in C: \\ j\neq i}} \abs{c_{C,j}} &\leq \sqrt{(\abs{C}-1)\sum_{\substack{j \in C: \\ j\neq i}} \abs{c_{C,j}}^2} \\
    &= \sqrt{(\abs{C}-1)} \sqrt{1- \abs{c_{C,i}}^2},
\end{aligned}
\end{equation}
and then over $i$, using
\begin{equation}
\begin{aligned}
    &\sum_{i \in C} \sqrt{1- \abs{c_{C,i}}^2} \left(\sqrt{ \expval{ n_i}_0 \expval{ n_C}_0}+\abs{c_{C,i}}\right) \\
    &\leq \sqrt{\sum_{i \in C} \left(1- \abs{c_{C,i}}^2\right)} \left(\sqrt{\sum_{i \in C} \expval{ n_i}_0 \expval{ n_C}_0} + \sqrt{\sum_{i \in C} \abs{c_{C,i}}^2}\right) \\
    &= \sqrt{(\abs{C}-1)} \left(\sqrt{\sum_{i \in C} \expval{ n_i}_0 \expval{ n_C}_0}+1\right).
\end{aligned}
\end{equation}
Combining these results, we express the bound for the sum over unique cross correlator pairs as
\begin{equation}
\begin{aligned}
    \sum_\text{pairs} \abs{\frac{\expval{b_C b_i^\dagger b_j b_C^\dagger}_0}{\expval{b_C b_C^\dagger}_0}-\expval{b_i^\dagger b_j}_0} \leq B_C,
\end{aligned}
\end{equation}
where
{\small \begin{equation} \label{eq:B_C_nonloc}
\begin{aligned}
    &B_C = \Bigg[ \frac{\sqrt{\expval{\Delta n_C^2}_0}}{2(\expval{n_C}_0+1)} \\
    &\times \sum_{j}\bigg(\sum_{i\in C: i \neq j}\sqrt{\expval{n_i n_j}_0 + \expval{n_j}_0} + \sum_{i\notin C: i \neq j}\sqrt{\expval{n_i n_j}_0 + \expval{n_i}_0}\bigg)\Bigg] \\
    &+ \frac{\abs{C}-1}{2(\expval{n_C}_0+1)} \bigg( \sqrt{\sum_{i \in C}\expval{ n_i}_0 \expval{ n_C}_0} + 1\bigg).
\end{aligned}
\end{equation}}

For the complete mixture \eqref{eq:1PAsemilocal}, the triangle inequality now readily gives
\begin{equation}
\begin{aligned}
    &\sum_\text{pairs} \abs{\expval{b_i^\dagger b_j}_{c}-\expval{b_i^\dagger b_j}_0} \\
    &\leq \sum_{C \in S} P_C \sum_\text{pairs} \abs{\frac{\expval{b_C b_i^\dagger b_j b_C^\dagger}_0}{\expval{b_C b_C^\dagger}_0}-\expval{b_i^\dagger b_j}_0}\leq \sum_{C \in S} P_C B_C.
\end{aligned}
\end{equation}
Nonlocality of at least $M+1$ simultaneous modes is witnessed upon violation of the above bound for all structures $S$ such that the number of modes in a combination $\abs{C}\leq M$ for all combinations in the structure. We therefore write the final bound as
\begin{equation} \label{eq:nonlocbound_distinguishM}
\begin{aligned}
    \sum_\text{pairs} &\abs{\expval{b_i^\dagger b_j}_{c}-\expval{b_i^\dagger b_j}_0} \leq \max_{\substack{S: \abs{C} \leq M \\
    \forall C \in S}} \left\{ \sum_{C\in S} P_C B_{C}\right\}.
\end{aligned}
\end{equation}
The probabilities $P_C$, given by \eqref{eq:general_loc_probs}, can be determined assuming measurability of the amplitudes $\abs{c_i}^2$ and the number statistics $\Tr{b_C^\dagger \rho_0 b_C} = \expval{n_C}+1$ of the truncated collective modes $b_C$. We note that since the witness requires measurement of number statistics $\expval{n_C}$ and $\expval{\Delta n_C^2}$ for every possible combination $C$ satisfying $\abs{C}\leq M$, the complexity of the measurement increases combinatorially for higher $N$. In Sec. \ref{sec:measurement_nonloc} we provide further details of the required measurements for Raman-scattering systems, for which the amplitudes $\abs{c_i}^2$ are independently measurable, and the statistics of the $C$-modes can be measured by switching off the drive signals corresponding to modes not in $C$.

\subsection{Example: Thermal \texorpdfstring{$W$}{W}-like state} \label{sec:nonlocwitness_therm_example}
We now consider the case of $N$ harmonic oscillators in a thermal initial state $\rho_0 = \rho_\text{th.}$, as in \eqref{eq:ThermalState}, and a conditioned state $\rho_c$, as in \eqref{eq:1PAgeneral}, with coefficients of equal magnitude $\abs{c_i}=1/\sqrt{N}$. The state $\rho_c$ does not conform to the assumption \eqref{eq:1PAsemilocal}, since the excitation is added nonlocally over all $N$ modes. Thermal states, which are symmetric with regard to all choices of modes, can be simply evaluated by setting $\expval{\Delta n_C^2}_\text{th.} = \nth(\nth+1)$ and $\expval{n_C}_\text{th.}=\nth$ for all combinations. Since the expressions are otherwise symmetric, the maximum in \eqref{eq:nonlocbound_distinguishM} is obtained by choosing a structure $S$ of combinations satisfying $\abs{C} \leq M$ which maximizes the number of commutator terms, i.e., the total number of times two modes fall within the same combination. This is accomplished by choosing $k$ disjoint combinations of size $M$, until there are less than $M$ modes remaining, and the final $N-kM$ modes as a remainder combination. The value of $k$ can be calculated via the integer division $k=\lfloor \frac{N}{M} \rfloor$. With the expectation values for thermal initial states substituted into \eqref{eq:nonlocbound_distinguishM}, we obtain the inequality
\begin{equation} \label{eq:nonlocbound_distinguishM_thermal}
\begin{aligned}
    &\frac{N(N-1)}{2} \frac{\nth+1}{N} \\
    &\leq \frac{N(N-1)}{2} \nth + \frac{kM^{3/2}(M-1)}{2N(\nth+1)} \left(\nth + \frac{1}{\sqrt{M}}\right) \\
    &+ \frac{(N-kM)^{3/2}(N-kM-1)}{2N(\nth+1)} \left(\nth + \frac{1}{\sqrt{N-kM}}\right).
\end{aligned}
\end{equation}
The threshold values of $\nth$ for which \eqref{eq:nonlocbound_distinguishM_thermal} is saturated are shown for various choices of $N$, $M$ in \tref{tab:relphys_thresholds}.

\begin{figure}[tb]
	% \begin{tikzpicture}[scale=1]
	% \begin{semilogyaxis}[xlabel=$N$, ylabel={$\nth$}, ymax=1, xticklabel style={anchor=north}, xmin=1.99, xmax = 10, xtick={2,3,4,5,6,7,8,9,10}, grid=both]
 %    \addplot+[color=blue] table [x index=0,y index=1,col sep=comma] {Plots/thresh_relwitness_max_new_ferm.csv};
 %    \addplot+[color=red] table [x index=0,y index=1,col sep=comma] {Plots/thresh_relwitness_max_new.csv};
	% \end{semilogyaxis}
	% \end{tikzpicture}
    \includegraphics{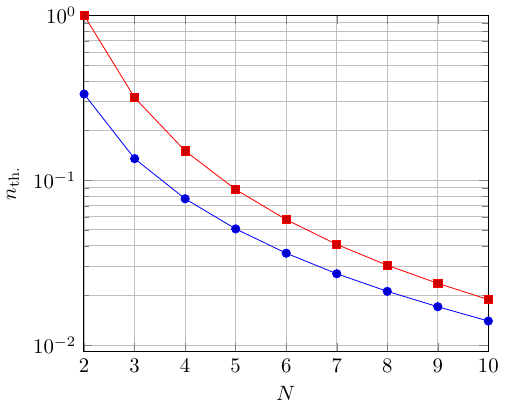}
	\caption{\label{fig:thresholds_relwitness}
	    Thermal occupation numbers $\nth$ which saturate the nonlocality witness \eqref{eq:nonlocbound_distinguishM_thermal} (in red) for $N$ harmonic oscillators in a thermal initial state, evaluated for $M=N-1$, corresponding to genuine $N$-partite nonlocality. The corresponding result for two-level systems is shown in blue. As opposed to the general entanglement witness (see \fref{fig:thresholds_abswitness_true}), where the harmonic oscillator and two-level system values quickly converge, here the two variants converge much slower. This is because adding a single excitation to a two-level system cannot occur if that system is already excited; an explicit dependence on the structure of the underlying Hilbert space is introduced with the assumption of a changing state. For the entanglement witness, on the other hand, only the proportion of the state which exceeds the first two levels (for harmonic oscillators) counts towards the difference, and quickly becomes negligible with decreasing $\nth$.
		}
\end{figure}

For a two-level system, the same argument can be followed, aside from the following modifications:
\begin{equation}
\begin{aligned}
    \expval{b_C b_C^\dagger}_0 &= 1-\expval{n_C}_0, \\
    \expval{b_j^\dagger b_i b_i^\dagger b_j}_0 &= \expval{n_j}_0-\expval{n_in_j}_0, \\
    \abs{\expval{b_i^\dagger b_j}_c} &=\abs{\expval{b_C b_i^\dagger b_j b_C^\dagger}_\text{th.}\big/\expval{b_C b_C^\dagger}_\text{th.}}= (1-\nth)/N, \\
    \expval{\Delta n_C^2}_\text{th.} &= \nth(1-\nth).
\end{aligned}
\end{equation}
The two-level equivalent of \eqref{eq:nonlocbound_distinguishM_thermal} is
\begin{equation} \label{eq:nonlocbound_distinguishM_thermal_2level}
\begin{aligned}
    &\frac{N(N-1)}{2} \frac{1-\nth}{N} \\
    &\leq \frac{N(N-1)}{2} \nth + \frac{kM^{3/2}(M-1)}{2N(1-\nth)} \left(\nth + \frac{1}{\sqrt{M}}\right) \\
    &+ \frac{(N-kM)^{3/2}(N-kM-1)}{2N(1-\nth)} \left(\nth + \frac{1}{\sqrt{N-kM}}\right).
\end{aligned}
\end{equation}
Thermal thresholds for witnessing genuine $N$-partite nonlocality (i.e., $M=N-1$) for both oscillators and two-level systems are given in \fref{fig:thresholds_relwitness}.

\begin{figure}[tb]
% 	\begin{tikzpicture}[scale=1]
% 	\begin{semilogxaxis}[xlabel=$\nth$, ylabel={$M+1$}, ymin = 1.7, ymax = 19, xticklabel style={anchor=north}, xmin=3e-3, xmax = 0.5, ytick={0,1,...,30}, grid=both, legend image post style={scale=0.5}, legend style={font=\footnotesize}]
%     \legend{TLS:$N=N_\text{opt.}(M)$, HO:$N=N_\text{opt.}(M)$, TLS:$N=M+1$, HO:$N=M+1$};
%     \addplot[color=blue, thick] table [x index=0,y index=1,col sep=comma] {Plots/relNoptfermvals.csv};
%     \addplot[color=red, thick] table [x index=0,y index=1,col sep=comma] {Plots/relNoptvals.csv};
%     \addplot[color=blue, dashed, thick] table [x index=0,y index=1,col sep=comma] {Plots/relNfermvals.csv};
%     \addplot[color=red, dashed, thick] table [x index=0,y index=1,col sep=comma] {Plots/relNvals.csv};

%     \addplot[draw=none, point meta =explicit,
% nodes near coords, every node near coord/.style={black!50, yshift=0pt, xshift=5pt}] table [x index=1,y index=0, meta index =2, col sep=comma] {Plots/thresh_relwitness_max_optN.csv};
% 	\end{semilogxaxis}
% 	\end{tikzpicture}
    \includegraphics{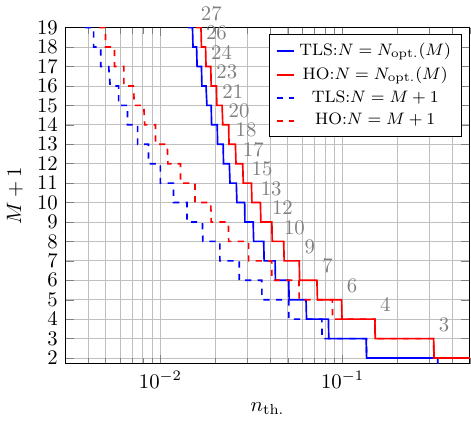}
	\caption{\label{fig:achievableMplus1_Nopt_nonlocwitness}
        The highest order of entanglement $M+1$ for which the inequalities (\eqref{eq:nonlocbound_distinguishM_thermal} in solid red, \eqref{eq:nonlocbound_distinguishM_thermal_2level} in solid blue) are violated for a thermal $W$-like state at a given thermal occupation $\nth$. The optimal number of subsystems $N_\text{opt.}$ is indicated by numbers above their corresponding horizontal grid lines, and are identical for the harmonic oscillator and two-level system cases. The values for $N=M+1$ are shown in dashed lines for comparison.
		}
\end{figure}

As was the case with the entanglement witness, non-monotonicity of the thresholds as functions of $N$ again implies the existence of some value $N=N_\text{opt.} \geq M+1$ which maximizes the thermal threshold required to witness nonlocality to the degree $M+1$. 
%One may therefore vary the number of subsystems to in order to maximize the degree of nonlocality which can be measured for a given achievable thermal occupation $\nth$.
\fref{fig:achievableMplus1_Nopt_nonlocwitness} shows the maximal achievable degree $M+1$ of nonlocality as a function of $\nth$, with the corresponding values of $N_\text{opt.}$ indicated. Again we refer to App. \ref{app:Opt_N}, where the near-linear relationship $N_\text{opt.}(M)$ is shown.

\begin{table}
    \centering
    \begin{tabular}{|p{20pt}|p{20pt}|p{20pt}|p{20pt}|p{20pt}|p{20pt}|p{20pt}|}
        \hline  &  M=1 & 2 & 3 & 4 & 5 & 6 \\\hline
        N=2 & 1.000 &  &  & & & \\ \hline
        3 & 0.500 & 0.317 & & & & \\ \hline
        4 & 0.333 & 0.214 & 0.151 & & & \\ \hline 
        5 & 0.250 & 0.197 & 0.142 & 0.088 &  & \\ \hline
        6 & 0.200 & 0.158 & 0.114 & 0.099 & 0.058 & \\ \hline
        7 & 0.167 & 0.142 & 0.115 & 0.090 & 0.073 & 0.041 \\ \hline
    \end{tabular}
    \caption{Thermal occupation numbers $\nth$ under which the nonlocality witness \eqref{eq:nonlocbound_distinguishM_thermal} for thermal initial states of $N$ harmonic oscillator systems is violated, as a function of the number of modes $N$ and the maximal number of simultaneously entangled modes $M$.}
    \label{tab:relphys_thresholds}
\end{table}

\section{Photodetective Measurements for Raman-Scattering Systems} \label{sec:measurement}
Our witnesses of genuine $M+1$-partite entanglement \eqref{eq:absNMineq} and nonlocality of the single excitation process \eqref{eq:nonlocbound_distinguishM} rely on the measurement of number statistics for the individual $b_i$-modes and their superpositions. This is naturally suited to the detection of Raman-scattered photons, for which linear combinations of modes can be accessed via networks of beam splitters. We will see below that this enables the measurement of all relevant expectation values for our witnesses \eqref{eq:absNMineq}, \eqref{eq:nonlocbound_distinguishM}. We take particular inspiration from optical control schemes in quantum optomechanics \cite{Aspelmeyer2014RMP, Vanner2013PRL, Bush2024PRA, Borkje2011PRL}, where single Raman-photon detection has been used to induce nonclassical behavior in mechanical oscillators \cite{Riedinger2016Nature, riedinger2018Nature, Hong2017Science, Velez2019PRX, Wang2023PhDThesis, Patel2021PRL, Enzian2021non}. However, we note that while we consider $b_i$ to be oscillator modes below, the analysis also applies to Raman-scattering from other degrees of freedom, e.g., atomic ensembles.

\subsection{Setup} \label{sec:setup}
We consider a system of $N$ identical, independent oscillators with resonance frequency $\omega_\text{osc.}$ and spectral linewidth $\gamma$. The oscillators, described via ladder operators $b_i$, are linearly coupled to corresponding optical cavity modes $a_i$. Incident coherent optical drive signals (i.e. laser light) of frequency $\Omega$ will, due to the linear coupling, cause energy to be transferred between the optical field and oscillators, producing Raman sidebands at frequencies $\Omega\pm\omega_\text{osc.}$. These processes correspond to excitations or de-excitations of the oscillators. Up-converted (anti-Stokes) photons at $\Omega+\omega_\text{osc.}$ are formed by removing quanta from the system. Conversely, down-converted (Stokes) photons at $\Omega-\omega_\text{osc.}$ correspond to adding quanta. This enables measurement of various oscillator correlations with equal numbers of creation and annihilation operators via photostatistics, as will be expanded on in Sec. \ref{sec:measurement_stokes_antistokes}, and can be used to produce heralded single-particle added (or removed) states via post selection.

\begin{figure}
\includegraphics{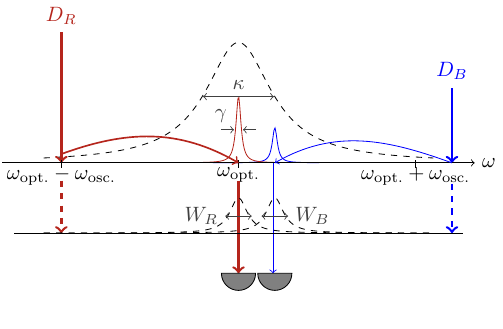}
\caption{\label{fig:Multitone_Scattering_Diagram}
       Diagram of the Raman-scattering processes of a coherently driven, linearly coupled oscillator-cavity system, with oscillator frequency $\omega_\text{osc.}$ and linewidth $\gamma$, and cavity resonance frequency $\omega_\text{opt.}$ and linewidth $\kappa$. Anti-Stokes photons from the red detuned ($D_{R}$) drive and Stokes photons from the blue detuned ($D_B$) drive can be separated by filters with linewidths $W_R, W_B \ll \omega_\text{osc.}$.
       }
\end{figure}

The measurement scheme we have in mind is illustrated in the scattering picture diagram \fref{fig:Multitone_Scattering_Diagram}. We consider the optical cavity modes to have resonance frequency $\omega_\text{opt.}$ and linewidth $\kappa$. Processes which lead to scattering into the cavity's resonant band will be enhanced, while all others will be suppressed. The ideal setup for accomplishing this is the resolved sideband regime,
\begin{align*}
    \kappa \ll \omega_\text{osc.},
\end{align*}
in which Raman-scattering out of the cavity resonance band is highly suppressed, while scattering into the band is enhanced. The limit 
%$\gamma \ll \kappa$
$\gamma_\mathrm{eff} \ll \kappa$
of large cavity linewidth relative to the {\it effective} oscillator decay rate ensures adiabatic response of the optical output to the oscillators. Additionally, it allows both Stokes and anti-Stokes photons to enter the cavity at mutually resolvable frequencies, if driving on both sidebands simultaneously. Appropriate filtering can then allow independent measurement of the respective sidebands, i.e., as independent Stokes and anti-Stokes optical modes $a_{S,i}$ and $a_{AS,i}$. Detailed derivations of the dynamics and properties of such systems can be found in \cite{Bush2024PRA, Borkje2011PRL}.

In the regime described above, one can show (see Sec. \ref{sec:measurement_stokes_antistokes}) that photodetection statistics of the Stokes and anti-Stokes photons, involving the operators $a_{S,i}^\dagger, a_{S,i}$ and $a_{AS,i}^\dagger, a_{AS,i}$, can be used to measure Hermitian expectation values involving an even number of the oscillator ladder operators, via the relations (up to vacuum noise terms)
\begin{subequations}\label{eq:SandASprops}
\begin{equation}
    a_{S,i} = \mathcal{G}_{S} b_i^\dagger
\end{equation}
and 
\begin{equation}
    a_{AS,i} = \mathcal{G}_{AS} b_i.
\end{equation}
\end{subequations}
The factors $\mathcal{G}_{S}$, $\mathcal{G}_{AS}$ represent the amplitudes of the sidebands along with detection gain.

\subsection{Linear Optical Networks and \texorpdfstring{$W$}{W}-State Preparation}
To access arbitrary superpositions of the optical, and, by extension, oscillator modes, one can use networks of beam splitters and phase shifts. The basic operation of a beam splitter is precisely to generate collective modes of its inputs,
\begin{equation}
\begin{aligned}
    a_{\text{out},A} &= \sqrt{r} a_{\text{in},1} + \sqrt{t} a_{\text{in},2}, \\
    a_{\text{out},B} &= \sqrt{t} a_{\text{in},1} - \sqrt{r} a_{\text{in},2},
\end{aligned}
\end{equation}
with reflection and transmission coefficients satisfying $r+t=1$. This makes such networks suitable for accessing nonlocal collective modes of the oscillators by measurement on a single output. Detecting a photon in a given output corresponds to removing a photon from a nonlocal, collective optical mode across its inputs, which in the case of a Raman-scattered photon means adding or removing excitations from the associated collective oscillator mode. The ability to make measurements on collective modes of the oscillators is crucial both in order to produce entangled single-excitation states, as well as to quantify their cross correlations. A network of several beam splitters and phase shifts can produce arbitrary linear combinations
\begin{equation}
    \begin{pmatrix}
        &a_1 \\
        &a_2 \\
        &a_3 \\
        &\vdots
    \end{pmatrix}
    \to
    \begin{pmatrix}
        &c_{11} a_1 + c_{12} a_2 + c_{13} a_3 + \ldots \\
        &c_{21} a_1 + c_{22} a_2 + c_{23} a_3 + \ldots \\
        &c_{31} a_1 + c_{32} a_2 + c_{33} a_3 + \ldots \\
        &\vdots
    \end{pmatrix}
\end{equation}
of the optical modes, under the constraint that the total optical flux is conserved in the absence of losses. A universal multiport interferometer can be used to implement any $N\times N$ unitary transformation like the above using at most $N(N-1)/2$ reprogrammable beam splitter elements \cite{Reck1994PRL, Clements2016optica}.

An equally weighted output mode
\begin{equation}
    a_W = \frac{1}{\sqrt{N}} (e^{i \theta_1}a_1 + e^{i \theta_2}a_2 + \ldots + e^{i \theta_N}a_N)
\end{equation}
will upon a Stokes photodetection produce an equally distributed single excitation across the oscillators. For harmonic oscillator systems in an initial thermal state \eqref{eq:ThermalState}, the measurement produces the thermal $W$-like state
\begin{equation} \label{eq:Wlikestate}
\begin{aligned}
    &\rho_c = \frac{(e^{-i \theta_1}b_1^\dagger + e^{-i \theta_2}b_2^\dagger + \ldots)\rho_\text{th.}(e^{i \theta_1}b_1 + e^{i \theta_2}b_2 + \ldots)}{N(\nth+1)}.
\end{aligned}
\end{equation}
As the thermal occupation number approaches zero and $\rho_\text{th.}\to\ket{0}\bra{0}$, it approximates the pure $W$ state
\begin{equation}
\begin{aligned}
    \ket{W_N} = \frac{1}{\sqrt{N}}(e^{-i \theta_1}\ket{1,0,\ldots} + e^{-i \theta_2}\ket{0,1,\ldots} + \ldots),
\end{aligned}
\end{equation}
as can be seen from the fidelity
\begin{equation}
    \bra{W_N}\rho_c\ket{W_N} = \frac{1}{(\nth + 1)^{N+1}}.
\end{equation}

\subsection{Measuring Oscillator Statistics Through Photodetection} \label{sec:measurement_stokes_antistokes}
It can be inferred from the relations \eqref{eq:SandASprops} that any expectation value of the form $\expval{B^\dagger B}$, where $B$ is some arbitrary product of oscillator ladder operators, may be represented as a normal ordered expectation value of the optical signals $a_{AS,i}$, $a_{S,i}$. This is done by performing substitutions in $B$ which result in a product of only optical annihilation operators, i.e., 
\begin{equation}
\begin{aligned}
    B &= (\ldots b_i \ldots b_j^\dagger \ldots) \\
    &= (\ldots a_{AS,i}/\mathcal{G}_{AS} \ldots a_{S,j}/\mathcal{G}_{S} \ldots).
\end{aligned}
\end{equation}
It follows that $B^\dagger$ will involve only optical creation operators, and $\expval{B^\dagger B}$ will be in normal order. Correlators of order $2$ (i.e., with $B$ being a product of two ladder operators) and above can naturally be measured via postselection, by using the state update rule
\begin{equation}
\begin{aligned}
    &\expval{O}_{c,S/AS,i} \equiv \frac{\expval{a_{S/AS,i}^\dagger O a_{S/AS,i}}}{\expval{a_{S/AS,i}^\dagger a_{S/AS,i}}} \\
    &\implies \expval{a_{S/AS,i}^\dagger O a_{S/AS,i}} = \expval{O}_{c,S/AS,i} \expval{a_{S/AS,i}^\dagger a_{S/AS,i}}
\end{aligned}
\end{equation}
for all operators $O$, where subscript $c,S/AS,i$ denotes conditioning on a Stokes or anti-Stokes photodetection in mode $i$.

\subsection{Measuring Simple Cross-Correlators} \label{sec:measurement_crosscorrelators}
Additional required expectation values are accessible by superposing the optical modes via linear optical networks. The simple cross correlator $\abs{\expval{b_1^\dagger b_2}}$, which is not directly measurable from photon counting, can be inferred from the statistics of the output modes $a_A, a_B$ of a $50:50$ beam splitter,
\begin{equation}
\begin{aligned}
    a_A &= \frac{1}{\sqrt{2}}(a_1+ e^{i\phi_2}a_2), \\
    a_B &= \frac{1}{\sqrt{2}}(a_1- e^{i\phi_2}a_2),
\end{aligned}
\end{equation}
where the phase of input $2$ is controlled by, e.g., extending the optical path length. Defining their corresponding collective oscillator modes
\begin{equation}
\begin{aligned}
    b_A &= \frac{1}{\sqrt{2}}(b_1+ e^{i\phi_2}b_2), \\
    b_B &= \frac{1}{\sqrt{2}}(b_1- e^{i\phi_2}b_2),
\end{aligned}
\end{equation}
one can show that
\begin{equation} \label{eq:simplebimodalcrosscorr}
    \cos (\theta + \phi_2) \abs{\expval{b_1^\dagger b_2}} = \frac{\expval{n_A}-\expval{n_B}}{2},
\end{equation}
where
\begin{equation}
\begin{aligned}
    \expval{n_A} &\equiv \expval{b_A^\dagger b_A} = \frac{\expval{a_{AS, A}^\dagger a_{AS, A}}}{\mathcal{G}_{AS}^2}, \\
    \expval{n_B} &\equiv \expval{b_B^\dagger b_B} = \frac{\expval{a_{AS, B}^\dagger a_{AS, B}}}{\mathcal{G}_{AS}^2}.
\end{aligned}
\end{equation}
In the absence of knowledge about the phase $\theta = \arg{\expval{b_1^\dagger b_2}}$, \eqref{eq:simplebimodalcrosscorr} can be maximized with respect to the variable phase $\phi_2$.

In the case of $N$-partite systems, our witnesses \eqref{eq:absNMineq} and \eqref{eq:nonlocbound_distinguishM} employ the sum of cross correlators over every unique pair of modes. Although this can be done as explained above for every pair, we can instead use the mode $a_W=\frac{1}{\sqrt{N}}\sum_j e^{i\phi_j}a_j$ to measure the entire sum at once by using the relation
\begin{equation}
    \sum_\text{pairs} \abs{\expval{b_i^\dagger b_j}} = \frac{N\max_{\{ \phi_j\}} \big\{ \expval{n_W}\big\} -\sum_j \expval{n_j}}{2},
\end{equation}
denoting maximization over the set of all relative phases, and where $n_W = b_W^\dagger b_W$ is the number operator of the corresponding oscillator mode. This approach requires access to only the individual modes $a_i$ and the optical mode $a_W$ which is required to generate the $W$-like state \eqref{eq:Wlikestate}, thereby imposing no additional constraint on the setup. Our nonlocality witness \eqref{eq:nonlocbound_distinguishM}, which is based on the differences in cross correlations $\abs{\expval{b_i^\dagger b_j}_c-\expval{b_i^\dagger b_j}_0}$ before and after state preparation, can similarly be recovered by the appropriate modification
\begin{equation}
\begin{aligned}
    \sum_\text{pairs} &\abs{\expval{b_i^\dagger b_j}_c-\expval{b_i^\dagger b_j}_0} \\
    &= \frac{N\max_{\{ \phi_j\}} \big\{ \expval{n_W}_c-\expval{n_W}_0\big\} -\sum_j (\expval{n_j}_c-\expval{n_j}_0)}{2}.
\end{aligned}
\end{equation}

\begin{figure}
    \includegraphics{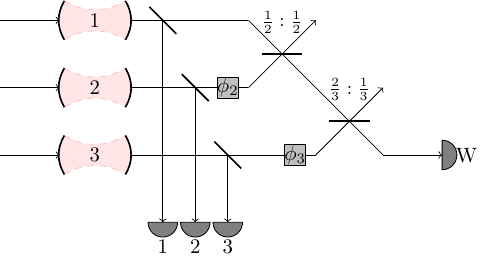}
        \caption{ \label{fig:setupdiagram}
            A possible tripartite setup which facilitates both the generation of a $W$-like state in the harmonic oscillator modes through Stokes sideband photodetection, as well as the measurement of all expectation values required by our witnesses \eqref{eq:absNMineq}, \eqref{eq:nonlocbound_distinguishM} through photodetection on both sidebands. The reflections and transmissions of the three unmarked beam splitters should be identical, but are otherwise arbitrary.
            }
\end{figure}

\fref{fig:setupdiagram} illustrates a setup which would facilitate these measurements for a tripartite system.

% Tripartite setup figure
\noindent

\subsection{Measurement of Entanglement Witness}
The two sides of the inequality \eqref{eq:absNMineq} can be measured purely from anti-Stokes photodetections, following the discussion in Secs. \ref{sec:measurement_stokes_antistokes} and \ref{sec:measurement_crosscorrelators}. With 
\begin{equation}
    \expval{n_in_j} = \frac{\expval{a_{AS,j}^\dagger a_{AS,i}^\dagger a_{AS,i} a_{AS,j}}}{\mathcal{G}_{AS}^4},
\end{equation}
the full witness is expressed as
\begin{equation} \label{eq:absNMineq_optical}
\begin{aligned}
    &\frac{N \max_{\{\phi_j\}} \left\{ \expval{a_{AS,W}^\dagger a_{AS,W}}\right\}-\sum_i\expval{a_{AS,i}^\dagger a_{AS,i}}}{2} \\ &\leq \frac{M-1}{2} \sum_{i=1}^N \expval{a_{AS,i}^\dagger a_{AS,i}} \\
    &+ \max_{i\neq j} \left\{\sqrt{\expval{a_{AS,j}^\dagger a_{AS,i}^\dagger a_{AS,i} a_{AS,j}}} \right\} \sqrt{N_\text{pairs} N_{\text{sep.},\mathcal{C}, \text{max}}}.
\end{aligned}
\end{equation}
The factor $1/\mathcal{G}_{AS}^2$ occurring in each term has been canceled out, meaning that the method is independent of optical gains and losses.

\subsection{Measurement of Nonlocality Witness} \label{sec:measurement_nonloc}
Whereas the expectation values in \eqref{eq:nonlocbound_distinguishM} involving the individual modes $b_i$ can be measured readily according to the preceding discussion, the remaining expectation values depend on the collective modes $b_C$. These modes are truncations of the mode $b_W$ used to generate the state \eqref{eq:1PAgeneral}, and their nonzero coefficients have the same relative phases and magnitudes as their corresponding coefficients in $b_W$. For a system in which the measurements are performed via photodetection of Raman-scattered photons, these modes can be directly accessed by disabling the input (or output) optical signals for the modes not in $C$, i.e., by setting $\abs{c_i}^2=0$ for $i \notin C$. For example, in the case of a tripartite system, one can access the mode $c_1 b_1+c_2 b_2$ by disabling the optical input for mode $3$. The statistics can then be read on the same output from which $b_W$ was accessed, requiring no change in the configuration of the setup. 

The expectation values which occur in the witness can be expressed in terms of the Raman-photon modes as
\begin{align}
    \expval{n_i n_j}_0 + \expval{n_i}_0 &= \expval{b_j b_i^\dagger b_i b_j^\dagger}_0 = \frac{\expval{a_{AS,i}^\dagger a_{S,j}^\dagger a_{S,j} a_{AS,i}}_0}{\mathcal{G}_{S}^2\mathcal{G}_{AS}^2}, \\
    \expval{n_C}_0+1 &= \expval{b_C b_C^\dagger}_0 =\frac{\expval{a_{S,C}^\dagger a_{S,C}}_0}{\mathcal{G}_{S}^2},
\end{align}
and
\begin{equation}
\begin{aligned}
    \expval{\Delta n_C^2}_0 &= \expval{b_C b_C^\dagger b_C b_C^\dagger}_0 - \expval{b_C b_C^\dagger}_0^2 \\
    &= \frac{\expval{a_{S,C}^\dagger a_{AS,C}^\dagger a_{AS,C} a_{S,C} }_0 - \lambda \expval{a_{S,C}^\dagger a_{S,C} }_0^2}{\mathcal{G}_{S}^2\mathcal{G}_{AS}^2},
\end{aligned}
\end{equation}
where $\lambda = \mathcal{G}_{AS}^2/\mathcal{G}_{S}^2$ is a dimensionless ratio of the gain parameters.

The first terms of \eqref{eq:B_C_nonloc} thereby take the form
\begin{equation}
\begin{aligned}
    &\frac{\sqrt{\expval{\Delta n_C^2}_0}}{\expval{n_C}_0+1} \sqrt{\expval{n_i n_j}_0 + \expval{n_j}_0}\\
    &= \sqrt{\expval{a_{S,C}^\dagger a_{AS,C}^\dagger a_{AS,C} a_{S,C} }_0 - \lambda \expval{a_{S,C}^\dagger a_{S,C} }_0^2} \\
    & \ \ \ \ \ \ \ \ \ \ \ \ \ \ \ \ \ \ \ \ \ \ \ \times\frac{\sqrt{\expval{a_{AS,i}^\dagger a_{S,j}^\dagger a_{S,j} a_{AS,i}}_0}}{\mathcal{G}_{AS}^2\expval{a_{S,C}^\dagger a_{S,C}}_0},
\end{aligned}
\end{equation}
and the last term is expressed as
\begin{equation} \label{eq:commterms_photo}
\begin{aligned}
    &\frac{\abs{C}-1}{2(\expval{n_C}_0+1)} \bigg( \sqrt{\sum_{i \in C}\expval{ n_i}_0 \expval{ n_C}_0} + 1\bigg) \\
    &=\frac{(\abs{C}-1)\mathcal{G}_{S}^2}{2\expval{a_{S,C}^\dagger a_{S,C}}_0\mathcal{G}_{AS}^2} \\
    &\times \bigg( \sqrt{\sum_{i \in C}\expval{ a_{AS,i}^\dagger a_{AS,i}}_0 \expval{ a_{AS,C}^\dagger a_{AS,C}}_0} + \mathcal{G}_{AS}^2\bigg).
\end{aligned}
\end{equation}
The factor $1/\mathcal{G}_{AS}^2$ appears in all terms on either side of the inequality, and can be canceled out. The final term does, however, contain a remaining dependence on the gain parameters. This was caused by the application of commutator relations in the derivation, which lead to a reduction of the number of ladder operator in the resulting terms. We can compensate for this by reintroducing commutators which equal $1$. Assuming knowledge of the ratio $\lambda$, the remaining gain factors can be substituted via the following identities for oscillator systems:
\begin{equation}
\begin{aligned}
    \mathcal{G}_{S}^2 &= \mathcal{G}_{S}^2 \expval{\comm{b_x}{b_x^\dagger}}_0 = \expval{a_{S,x}^\dagger a_{S,x}}_0-\frac{1}{\lambda} \expval{a_{AS,x}^\dagger a_{AS,x}}_0, \\
    \mathcal{G}_{AS}^2 &= \mathcal{G}_{AS}^2 \expval{\comm{b_x}{b_x^\dagger}}_0 = \lambda \expval{a_{S,x}^\dagger a_{S,x}}_0- \expval{a_{AS,x}^\dagger a_{AS,x}}_0,
\end{aligned}
\end{equation}
where $x$ is an arbitrary mode.

The probabilities $P_C$, defined in \eqref{eq:general_loc_probs}, also depend directly on the amplitudes $\abs{c_i}^2$ of the mode $b_W$. These can be measured by the photocounting rates of mode $b_{C_i}$, where $C_i=\{ i\}$ is a single mode combination, in an initial thermal state. They are given in normalized form by the ratio
\begin{equation}
    \frac{\expval{n_{C_i}}_\text{th.}}{\expval{n_W}_\text{th.}} = \frac{\abs{c_i}^2\expval{n_i}_\text{th.}}{\sum_j \abs{c_j}^2\expval{n_j}_\text{th.}} = \frac{\abs{c_i}^2}{\sum_j \abs{c_j}^2}.
\end{equation}
We note that in order to accurately represent the probabilities of photons arriving at the detector of mode $b_W$ from a given local mode $i$, the modes $b_{C_i}$ must be measured on the same output as the mode $b_W$, in the way described in the beginning of this section.

\section{Discussion} \label{sec:discussion}
Our witnesses for genuine multipartite entanglement \eqref{eq:absNMineq} and nonlocality of the single excitation process \eqref{eq:nonlocbound_distinguishM} are derived with two main considerations in mind. Firstly, they should be general, in the sense that they are applicable to a wide range of possible systems. We have used the case of Raman scattering harmonic oscillator systems as the most relevant example. Secondly, they should depend on the system's readily accessible observables (in this case number statistics), and not rely on state tomography.

The entanglement witness \eqref{eq:absNMineq} detects the entanglement of $W$-states for an arbitrary number of modes, and requires only number expectation values of the modes up to second order, making it particularly accessible. One should note, however, that violation of the inequality necessitates nonzero bimodal cross-correlators $\abs{\expval{b_i^\dagger b_j}}$, and it can therefore never detect the entanglement of, e.g., Greenberger–Horne–Zeilinger (GHZ) states or multimode squeezed vacuum states, for which such expectation values are zero. Deriving a similar witness for GHZ states could possibly be done based on the higher order correlator $\abs{\expval{b_1^\dagger b_2^\dagger \ldots b_N^\dagger}}$, which, as with GHZ state preparation, is not directly accessible through photodetection and is therefore beyond the scope of this article.

The nonlocality witness \eqref{eq:nonlocbound_distinguishM} may also be used to witness entanglement of the final state, whenever the initial state can be assumed to be separable. In this case, nonlocality of the single excitation process always implies that the resulting state is entangled, as seen from the fact that for a fully separable initial state,
\begin{equation}
    b_W^\dagger \rho_\text{sep.} b_W \equiv \sum_k P_k \sum_{ij} c_i^* c_j b_i^\dagger \rho_{1,k} \otimes \ldots \otimes \rho_{N,k} b_j
\end{equation}
always contains inseparable intermodal terms. In other words, if the nonlocality witness is violated, it means that either the resulting state \emph{is} entangled, the initial state \emph{was} entangled (and belonged to a sparse set of states whose entanglement was precisely undone by the process), or both. We consider the assumption of initial separability to be reasonable for remote systems, which tend to quickly equilibrate to thermal steady states when left alone. This assumption lets our nonlocality witness act as an alternative entanglement witness for states such as the $W$-like states \eqref{eq:NpartiteWlikestate}, with less strict requirements on the initial thermal occupation. Detecting $M+1$-partite nonlocality corresponds in this case precisely to $M+1$-partite entanglement as in our entanglement witness.

\begin{figure*}[th]

    \includegraphics{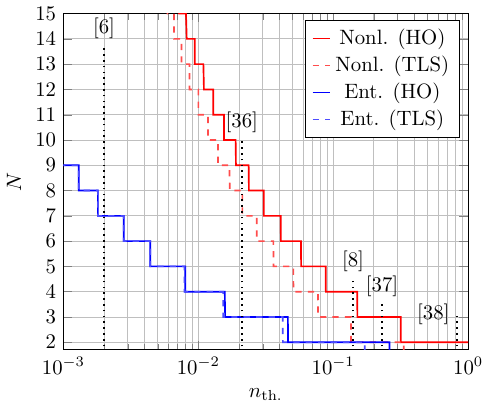}%
    \includegraphics{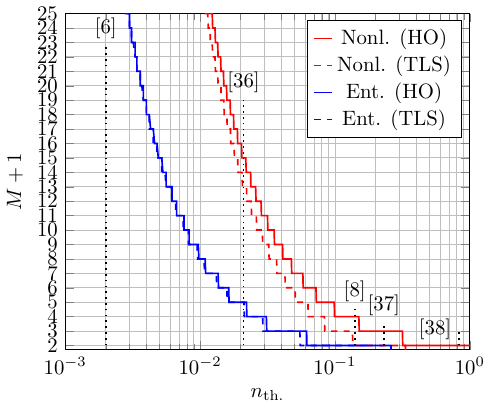}
	\caption{\label{fig:achievableN}
	    Left: The largest number of subsystems $N$ for which our witnesses are capable of verifying genuine entanglement (i.e., $M = N-1$) for thermal $W$-like states (\eqref{eq:threshold_sym_thermal} and \eqref{eq:threshold_sym_thermal_2level} in blue) and nonlocality (\eqref{eq:nonlocbound_distinguishM_thermal} and \eqref{eq:nonlocbound_distinguishM_thermal_2level} in red) as functions of initial thermal occupation $\nth$. The results for harmonic oscillators (HO) are shown in solid lines while the corresponding results for two-level systems (TLS) are dashed. Some thermal occupation numbers achieved in relevant experiments are indicated by dotted lines: Optomechanical crystals ($\nth\approx0.021$ \cite{Meenehan2015PRX} and $\nth\approx0.14$ \cite{riedinger2018Nature}), membrane flexural modes ($\nth\approx0.23$ \cite{Galinskiy2020Optica}), acoustic modes in confined superfluid helium ($\nth\approx0.83$ \cite{Patil2022PRL}), and $40$THz vibrational modes of diamond crystal lattices ($\nth\approx 0.002$ at room temperature \cite{Lee2011Science}, for which nonlocality can be witnessed between $N=30$ harmonic oscillators). Experiments with atomic ensembles are not shown as they are unlikely to be limited by thermal occupation. 
        Right: The maximal degree of entanglement/nonlocality $M+1$ achievable for a given thermal occupation number $\nth$ assuming an optimal number of subsystems $N=N_\text{opt.}\geq M+1$. 
		}
\end{figure*}

The witnesses we have derived are capable of detecting entanglement and nonlocality, respectively, for thermal $W$-like states \eqref{eq:NpartiteWlikestate} with thermal occupation numbers $\nth$ 
%within experimentally achievable limits 
below particular threshold values (see Tables \ref{tab:sym_thresholds}, \ref{tab:relphys_thresholds} and Figures \ref{fig:thresholds_abswitness_true}, \ref{fig:thresholds_relwitness}). For a given temperature, the value of $\nth$ will depend on the type of system in question, in particular the excitation energy of the modes and the applied methods of cooling. The leftmost plot in \fref{fig:achievableN} shows the maximal number of modes $N$ for which the witnesses with $M=N-1$ are violated as functions of $\nth$, where values of $\nth$ realized in some relevant experimental setups are also indicated.  As can be seen from the plot, the advantage of using nonlocality as a substitute entanglement witness for thermal $W$-like states is considerable for lower thermal occupations. Additionally, the rightmost plot in \fref{fig:achievableN} shows the maximal degree of observable entanglement/nonlocality where, rather than assuming $N=M+1$, for each degree of entanglement or nonlocality $M+1$ the number of subsystems is allowed to take the optimal value $N_\text{opt.}$ that maximizes the thermal threshold (as in \fref{fig:achievableMplus1_Nopt_entwitness}, \fref{fig:achievableMplus1_Nopt_nonlocwitness}). It is clear that if no limit is placed on the number of subsystems, this approach allows one to witness significantly higher degrees of entanglement and nonlocality for a fixed value of $\nth$. The optimal number of subsystems $N_\text{opt.}$ is roughly linear as a function of $M$, as seen in App. \ref{app:Opt_N}.

We have throughout our derivations assumed that the subsystems in question are indistinguishable; if at any point information about which state a given subsystem occupies is determined (for instance, through interaction with the environment or part of the instrumentation), the entanglement is broken. In our case this means in particular that it should be impossible to determine from which system a given optical signal comes from, aside from its location, i.e., they should produce output sidebands which are as close to identical in frequency and linewidth as possible. Indistinguishability is a natural feature for systems such as atomic ensembles, but is more challenging to achieve for engineered systems such as mechanical oscillators and optical cavities. To achieve sufficient indistinguishability across several modes, the possibility of \emph{in-situ} tuning may be advantageous. One promising optomechanical platform is superfluid helium-filled fiber cavities \cite{Patil2022PRL}, where the acoustic mode frequencies could be tuned by adjusting the cavity lengths.

In previous literature on entanglement witnesses for two-level systems, noise is often modeled in terms of a contribution $p I_N/2^N$ of white noise added to the pure state of interest, where $I_N$ is the identity operator and $N$ is the number of subsystems \cite{Guhne2010NJP, guhne2007toolbox, zhao2019efficient, zhou2020entanglement}. This is clearly not a reasonable model for noise in continuous variable systems, where white noise contributions are non-normalizable. We have therefore chosen to use initial thermal occupation as the main measure of robustness in this article. However, for the sake of comparison, for the tripartite two-level white noise case
\begin{equation}
    \rho = (1-p)\ket{W_3}\bra{W_3} + pI_3/8,
\end{equation}
our entanglement witness detects genuine tripartite entanglement below the threshold $p \approx 0.127$. For the partial tomography-based witness in Ref. \cite{Guhne2010NJP} (their Eq. (4)), the threshold value is $p \approx 0.471$. This exemplifies a tradeoff between witnesses based on natural observables and those based on partial tomography. Of course, the limit case for partial tomography is a fully tomographic characterization of the entanglement. However, this advantage comes at the cost of a number of required measurements which increases rapidly with the number $N$ of subsystems. For two-level systems, the most efficient conventional scheme for detecting the entanglement of $W$-states requires $2N-1$ so-called \emph{measurement settings}, each corresponding to a local measurement of all $N$ subsystems in parallel \cite{guhne2007toolbox}. The cost of tomography is particularly notable for continuous variable systems, for which the number of measurements required increases exponentially with $N$, e.g., by measurement of a $2N$-dimensional Wigner quasiprobability distribution, which even with optimized schemes can require hundreds of distinct measurements for a single system ($N=1$) \cite{landon2018quantitative}. Our witness requires only one measurement setting with one additional measurement of a nonlocal mode, for any underlying Hilbert space dimension and all values of $N$.

With the above discussion in mind, we believe the results presented in this article are relevant for the next generation of experiments investigating entanglement of multipartite Raman-scattering systems.

\acknowledgments{We acknowledge financial support from the Research Council of Norway (Grant No. 333937) through participation in the QuantERA
ERA-NET Cofund in Quantum Technologies (project MQSENS).}

\FloatBarrier

%===== Bibliography==========
%\bibliographystyle{apsrev4-2}
%\bibliography{article3}

\begin{thebibliography}{42}%
\makeatletter
\providecommand \@ifxundefined [1]{%
 \@ifx{#1\undefined}
}%
\providecommand \@ifnum [1]{%
 \ifnum #1\expandafter \@firstoftwo
 \else \expandafter \@secondoftwo
 \fi
}%
\providecommand \@ifx [1]{%
 \ifx #1\expandafter \@firstoftwo
 \else \expandafter \@secondoftwo
 \fi
}%
\providecommand \natexlab [1]{#1}%
\providecommand \enquote  [1]{``#1''}%
\providecommand \bibnamefont  [1]{#1}%
\providecommand \bibfnamefont [1]{#1}%
\providecommand \citenamefont [1]{#1}%
\providecommand \href@noop [0]{\@secondoftwo}%
\providecommand \href [0]{\begingroup \@sanitize@url \@href}%
\providecommand \@href[1]{\@@startlink{#1}\@@href}%
\providecommand \@@href[1]{\endgroup#1\@@endlink}%
\providecommand \@sanitize@url [0]{\catcode `\\12\catcode `\$12\catcode `\&12\catcode `\#12\catcode `\^12\catcode `\_12\catcode `\%12\relax}%
\providecommand \@@startlink[1]{}%
\providecommand \@@endlink[0]{}%
\providecommand \url  [0]{\begingroup\@sanitize@url \@url }%
\providecommand \@url [1]{\endgroup\@href {#1}{\urlprefix }}%
\providecommand \urlprefix  [0]{URL }%
\providecommand \Eprint [0]{\href }%
\providecommand \doibase [0]{https://doi.org/}%
\providecommand \selectlanguage [0]{\@gobble}%
\providecommand \bibinfo  [0]{\@secondoftwo}%
\providecommand \bibfield  [0]{\@secondoftwo}%
\providecommand \translation [1]{[#1]}%
\providecommand \BibitemOpen [0]{}%
\providecommand \bibitemStop [0]{}%
\providecommand \bibitemNoStop [0]{.\EOS\space}%
\providecommand \EOS [0]{\spacefactor3000\relax}%
\providecommand \BibitemShut  [1]{\csname bibitem#1\endcsname}%
\let\auto@bib@innerbib\@empty
%</preamble>
\bibitem [{\citenamefont {Bennett}\ \emph {et~al.}(1993)\citenamefont {Bennett}, \citenamefont {Brassard}, \citenamefont {Cr\'epeau}, \citenamefont {Jozsa}, \citenamefont {Peres},\ and\ \citenamefont {Wootters}}]{Bennett1993PRL}%
  \BibitemOpen
  \bibfield  {author} {\bibinfo {author} {\bibfnamefont {C.~H.}\ \bibnamefont {Bennett}}, \bibinfo {author} {\bibfnamefont {G.}~\bibnamefont {Brassard}}, \bibinfo {author} {\bibfnamefont {C.}~\bibnamefont {Cr\'epeau}}, \bibinfo {author} {\bibfnamefont {R.}~\bibnamefont {Jozsa}}, \bibinfo {author} {\bibfnamefont {A.}~\bibnamefont {Peres}},\ and\ \bibinfo {author} {\bibfnamefont {W.~K.}\ \bibnamefont {Wootters}},\ }\bibfield  {title} {\bibinfo {title} {Teleporting an unknown quantum state via dual classical and einstein-podolsky-rosen channels},\ }\href {https://doi.org/10.1103/PhysRevLett.70.1895} {\bibfield  {journal} {\bibinfo  {journal} {Phys. Rev. Lett.}\ }\textbf {\bibinfo {volume} {70}},\ \bibinfo {pages} {1895} (\bibinfo {year} {1993})}\BibitemShut {NoStop}%
\bibitem [{\citenamefont {Kruszynska}\ \emph {et~al.}(2006)\citenamefont {Kruszynska}, \citenamefont {Anders}, \citenamefont {D{\"u}r},\ and\ \citenamefont {Briegel}}]{kruszynska2006PRA}%
  \BibitemOpen
  \bibfield  {author} {\bibinfo {author} {\bibfnamefont {C.}~\bibnamefont {Kruszynska}}, \bibinfo {author} {\bibfnamefont {S.}~\bibnamefont {Anders}}, \bibinfo {author} {\bibfnamefont {W.}~\bibnamefont {D{\"u}r}},\ and\ \bibinfo {author} {\bibfnamefont {H.~J.}\ \bibnamefont {Briegel}},\ }\bibfield  {title} {\bibinfo {title} {Quantum communication cost of preparing multipartite entanglement},\ }\href@noop {} {\bibfield  {journal} {\bibinfo  {journal} {Physical Review A—Atomic, Molecular, and Optical Physics}\ }\textbf {\bibinfo {volume} {73}},\ \bibinfo {pages} {062328} (\bibinfo {year} {2006})}\BibitemShut {NoStop}%
\bibitem [{\citenamefont {Zhang}\ and\ \citenamefont {Zhuang}(2021)}]{zhang2021QST}%
  \BibitemOpen
  \bibfield  {author} {\bibinfo {author} {\bibfnamefont {Z.}~\bibnamefont {Zhang}}\ and\ \bibinfo {author} {\bibfnamefont {Q.}~\bibnamefont {Zhuang}},\ }\bibfield  {title} {\bibinfo {title} {Distributed quantum sensing},\ }\href@noop {} {\bibfield  {journal} {\bibinfo  {journal} {Quantum Science and Technology}\ }\textbf {\bibinfo {volume} {6}},\ \bibinfo {pages} {043001} (\bibinfo {year} {2021})}\BibitemShut {NoStop}%
\bibitem [{\citenamefont {Duan}\ \emph {et~al.}(2001)\citenamefont {Duan}, \citenamefont {Lukin}, \citenamefont {Cirac},\ and\ \citenamefont {Zoller}}]{Duan2001Nature}%
  \BibitemOpen
  \bibfield  {author} {\bibinfo {author} {\bibfnamefont {L.-M.}\ \bibnamefont {Duan}}, \bibinfo {author} {\bibfnamefont {M.~D.}\ \bibnamefont {Lukin}}, \bibinfo {author} {\bibfnamefont {J.~I.}\ \bibnamefont {Cirac}},\ and\ \bibinfo {author} {\bibfnamefont {P.}~\bibnamefont {Zoller}},\ }\bibfield  {title} {\bibinfo {title} {Long-distance quantum communication with atomic ensembles and linear optics},\ }\href@noop {} {\bibfield  {journal} {\bibinfo  {journal} {Nature}\ }\textbf {\bibinfo {volume} {414}},\ \bibinfo {pages} {413} (\bibinfo {year} {2001})}\BibitemShut {NoStop}%
\bibitem [{\citenamefont {Chou}\ \emph {et~al.}(2005)\citenamefont {Chou}, \citenamefont {de~Riedmatten}, \citenamefont {Felinto}, \citenamefont {Polyakov}, \citenamefont {van Enk},\ and\ \citenamefont {Kimble}}]{Chou2005Nature}%
  \BibitemOpen
  \bibfield  {author} {\bibinfo {author} {\bibfnamefont {C.~W.}\ \bibnamefont {Chou}}, \bibinfo {author} {\bibfnamefont {H.}~\bibnamefont {de~Riedmatten}}, \bibinfo {author} {\bibfnamefont {D.}~\bibnamefont {Felinto}}, \bibinfo {author} {\bibfnamefont {S.~V.}\ \bibnamefont {Polyakov}}, \bibinfo {author} {\bibfnamefont {S.~J.}\ \bibnamefont {van Enk}},\ and\ \bibinfo {author} {\bibfnamefont {H.~J.}\ \bibnamefont {Kimble}},\ }\bibfield  {title} {\bibinfo {title} {Measurement-induced entanglement for excitation stored in remote atomic ensembles},\ }\href@noop {} {\bibfield  {journal} {\bibinfo  {journal} {Nature}\ }\textbf {\bibinfo {volume} {438}},\ \bibinfo {pages} {828} (\bibinfo {year} {2005})}\BibitemShut {NoStop}%
\bibitem [{\citenamefont {Lee}\ \emph {et~al.}(2011)\citenamefont {Lee}, \citenamefont {Sprague}, \citenamefont {Sussman}, \citenamefont {Nunn}, \citenamefont {Langford}, \citenamefont {Jin}, \citenamefont {Champion}, \citenamefont {Michelberger}, \citenamefont {Reim}, \citenamefont {England}, \citenamefont {Jaksch},\ and\ \citenamefont {Walmsley}}]{Lee2011Science}%
  \BibitemOpen
  \bibfield  {author} {\bibinfo {author} {\bibfnamefont {K.~C.}\ \bibnamefont {Lee}}, \bibinfo {author} {\bibfnamefont {M.~R.}\ \bibnamefont {Sprague}}, \bibinfo {author} {\bibfnamefont {B.~J.}\ \bibnamefont {Sussman}}, \bibinfo {author} {\bibfnamefont {J.}~\bibnamefont {Nunn}}, \bibinfo {author} {\bibfnamefont {N.~K.}\ \bibnamefont {Langford}}, \bibinfo {author} {\bibfnamefont {X.-M.}\ \bibnamefont {Jin}}, \bibinfo {author} {\bibfnamefont {T.}~\bibnamefont {Champion}}, \bibinfo {author} {\bibfnamefont {P.}~\bibnamefont {Michelberger}}, \bibinfo {author} {\bibfnamefont {K.~F.}\ \bibnamefont {Reim}}, \bibinfo {author} {\bibfnamefont {D.}~\bibnamefont {England}}, \bibinfo {author} {\bibfnamefont {D.}~\bibnamefont {Jaksch}},\ and\ \bibinfo {author} {\bibfnamefont {I.~A.}\ \bibnamefont {Walmsley}},\ }\bibfield  {title} {\bibinfo {title} {Entangling macroscopic diamonds at room temperature},\ }\href {https://doi.org/10.1126/science.1211914} {\bibfield  {journal} {\bibinfo  {journal} {Science}\ }\textbf {\bibinfo
  {volume} {334}},\ \bibinfo {pages} {1253} (\bibinfo {year} {2011})}\BibitemShut {NoStop}%
\bibitem [{\citenamefont {Usmani}\ \emph {et~al.}(2012)\citenamefont {Usmani}, \citenamefont {Clausen}, \citenamefont {Bussi{\`e}res}, \citenamefont {Sangouard}, \citenamefont {Afzelius},\ and\ \citenamefont {Gisin}}]{Usmani2012NatPhys}%
  \BibitemOpen
  \bibfield  {author} {\bibinfo {author} {\bibfnamefont {I.}~\bibnamefont {Usmani}}, \bibinfo {author} {\bibfnamefont {C.}~\bibnamefont {Clausen}}, \bibinfo {author} {\bibfnamefont {F.}~\bibnamefont {Bussi{\`e}res}}, \bibinfo {author} {\bibfnamefont {N.}~\bibnamefont {Sangouard}}, \bibinfo {author} {\bibfnamefont {M.}~\bibnamefont {Afzelius}},\ and\ \bibinfo {author} {\bibfnamefont {N.}~\bibnamefont {Gisin}},\ }\bibfield  {title} {\bibinfo {title} {Heralded quantum entanglement between two crystals},\ }\href@noop {} {\bibfield  {journal} {\bibinfo  {journal} {Nature Photonics}\ }\textbf {\bibinfo {volume} {6}},\ \bibinfo {pages} {234} (\bibinfo {year} {2012})}\BibitemShut {NoStop}%
\bibitem [{\citenamefont {Riedinger}\ \emph {et~al.}(2018)\citenamefont {Riedinger}, \citenamefont {Wallucks}, \citenamefont {Marinkovi{\'c}}, \citenamefont {L{\"o}schnauer}, \citenamefont {Aspelmeyer}, \citenamefont {Hong},\ and\ \citenamefont {Gr{\"o}blacher}}]{riedinger2018Nature}%
  \BibitemOpen
  \bibfield  {author} {\bibinfo {author} {\bibfnamefont {R.}~\bibnamefont {Riedinger}}, \bibinfo {author} {\bibfnamefont {A.}~\bibnamefont {Wallucks}}, \bibinfo {author} {\bibfnamefont {I.}~\bibnamefont {Marinkovi{\'c}}}, \bibinfo {author} {\bibfnamefont {C.}~\bibnamefont {L{\"o}schnauer}}, \bibinfo {author} {\bibfnamefont {M.}~\bibnamefont {Aspelmeyer}}, \bibinfo {author} {\bibfnamefont {S.}~\bibnamefont {Hong}},\ and\ \bibinfo {author} {\bibfnamefont {S.}~\bibnamefont {Gr{\"o}blacher}},\ }\bibfield  {title} {\bibinfo {title} {Remote quantum entanglement between two micromechanical oscillators},\ }\href@noop {} {\bibfield  {journal} {\bibinfo  {journal} {Nature}\ }\textbf {\bibinfo {volume} {556}},\ \bibinfo {pages} {473} (\bibinfo {year} {2018})}\BibitemShut {NoStop}%
\bibitem [{\citenamefont {Joo}\ \emph {et~al.}(2002)\citenamefont {Joo}, \citenamefont {Lee}, \citenamefont {Jang},\ and\ \citenamefont {Park}}]{Joo2002}%
  \BibitemOpen
  \bibfield  {author} {\bibinfo {author} {\bibfnamefont {J.}~\bibnamefont {Joo}}, \bibinfo {author} {\bibfnamefont {J.}~\bibnamefont {Lee}}, \bibinfo {author} {\bibfnamefont {J.}~\bibnamefont {Jang}},\ and\ \bibinfo {author} {\bibfnamefont {Y.-J.}\ \bibnamefont {Park}},\ }\bibfield  {title} {\bibinfo {title} {Quantum secure communication via w states}} (\bibinfo {year} {2002}),\ \bibinfo {note} {e-print quant-ph/0204003}\BibitemShut {NoStop}%
\bibitem [{\citenamefont {Joo}\ \emph {et~al.}(2003)\citenamefont {Joo}, \citenamefont {Park}, \citenamefont {Oh},\ and\ \citenamefont {Kim}}]{Joo2003NJP}%
  \BibitemOpen
  \bibfield  {author} {\bibinfo {author} {\bibfnamefont {J.}~\bibnamefont {Joo}}, \bibinfo {author} {\bibfnamefont {Y.-J.}\ \bibnamefont {Park}}, \bibinfo {author} {\bibfnamefont {S.}~\bibnamefont {Oh}},\ and\ \bibinfo {author} {\bibfnamefont {J.}~\bibnamefont {Kim}},\ }\bibfield  {title} {\bibinfo {title} {Quantum teleportation via a w state},\ }\href {https://doi.org/10.1088/1367-2630/5/1/136} {\bibfield  {journal} {\bibinfo  {journal} {New Journal of Physics}\ }\textbf {\bibinfo {volume} {5}},\ \bibinfo {pages} {136} (\bibinfo {year} {2003})}\BibitemShut {NoStop}%
\bibitem [{\citenamefont {Agrawal}\ and\ \citenamefont {Pati}(2006)}]{Agrawal2006PRA}%
  \BibitemOpen
  \bibfield  {author} {\bibinfo {author} {\bibfnamefont {P.}~\bibnamefont {Agrawal}}\ and\ \bibinfo {author} {\bibfnamefont {A.}~\bibnamefont {Pati}},\ }\bibfield  {title} {\bibinfo {title} {Perfect teleportation and superdense coding with $w$ states},\ }\href {https://doi.org/10.1103/PhysRevA.74.062320} {\bibfield  {journal} {\bibinfo  {journal} {Phys. Rev. A}\ }\textbf {\bibinfo {volume} {74}},\ \bibinfo {pages} {062320} (\bibinfo {year} {2006})}\BibitemShut {NoStop}%
\bibitem [{\citenamefont {Liu}\ \emph {et~al.}(2012)\citenamefont {Liu}, \citenamefont {Tsai},\ and\ \citenamefont {Hwang}}]{Liu2012IJTP}%
  \BibitemOpen
  \bibfield  {author} {\bibinfo {author} {\bibfnamefont {L.-L.}\ \bibnamefont {Liu}}, \bibinfo {author} {\bibfnamefont {C.-W.}\ \bibnamefont {Tsai}},\ and\ \bibinfo {author} {\bibfnamefont {T.}~\bibnamefont {Hwang}},\ }\bibfield  {title} {\bibinfo {title} {Quantum secret sharing using symmetric w state},\ }\href {https://doi.org/10.1007/s10773-012-1109-7} {\bibfield  {journal} {\bibinfo  {journal} {International Journal of Theoretical Physics}\ }\textbf {\bibinfo {volume} {51}},\ \bibinfo {pages} {2291} (\bibinfo {year} {2012})}\BibitemShut {NoStop}%
\bibitem [{\citenamefont {Li}\ \emph {et~al.}(2024)\citenamefont {Li}, \citenamefont {Cheng}, \citenamefont {Wang}, \citenamefont {Cheng}, \citenamefont {Mao},\ and\ \citenamefont {Jia}}]{li2024quantum}%
  \BibitemOpen
  \bibfield  {author} {\bibinfo {author} {\bibfnamefont {G.-D.}\ \bibnamefont {Li}}, \bibinfo {author} {\bibfnamefont {W.-C.}\ \bibnamefont {Cheng}}, \bibinfo {author} {\bibfnamefont {Q.-L.}\ \bibnamefont {Wang}}, \bibinfo {author} {\bibfnamefont {L.}~\bibnamefont {Cheng}}, \bibinfo {author} {\bibfnamefont {Y.}~\bibnamefont {Mao}},\ and\ \bibinfo {author} {\bibfnamefont {H.-Y.}\ \bibnamefont {Jia}},\ }\bibfield  {title} {\bibinfo {title} {Quantum secret sharing enhanced: Utilizing w states for anonymous and secure communication},\ }\href@noop {} {\bibfield  {journal} {\bibinfo  {journal} {arXiv preprint arXiv:2402.02413}\ } (\bibinfo {year} {2024})}\BibitemShut {NoStop}%
\bibitem [{\citenamefont {Zou}\ \emph {et~al.}(2002)\citenamefont {Zou}, \citenamefont {Pahlke},\ and\ \citenamefont {Mathis}}]{zou2002generation}%
  \BibitemOpen
  \bibfield  {author} {\bibinfo {author} {\bibfnamefont {X.}~\bibnamefont {Zou}}, \bibinfo {author} {\bibfnamefont {K.}~\bibnamefont {Pahlke}},\ and\ \bibinfo {author} {\bibfnamefont {W.}~\bibnamefont {Mathis}},\ }\bibfield  {title} {\bibinfo {title} {Generation of an entangled four-photon w state},\ }\href@noop {} {\bibfield  {journal} {\bibinfo  {journal} {Physical Review A}\ }\textbf {\bibinfo {volume} {66}},\ \bibinfo {pages} {044302} (\bibinfo {year} {2002})}\BibitemShut {NoStop}%
\bibitem [{\citenamefont {Kiesel}\ \emph {et~al.}(2003)\citenamefont {Kiesel}, \citenamefont {Bourennane}, \citenamefont {Kurtsiefer}, \citenamefont {Weinfurter}, \citenamefont {Kaszlikowski}, \citenamefont {Laskowski},\ and\ \citenamefont {Zukowski}}]{kiesel2003three}%
  \BibitemOpen
  \bibfield  {author} {\bibinfo {author} {\bibfnamefont {N.}~\bibnamefont {Kiesel}}, \bibinfo {author} {\bibfnamefont {M.}~\bibnamefont {Bourennane}}, \bibinfo {author} {\bibfnamefont {C.}~\bibnamefont {Kurtsiefer}}, \bibinfo {author} {\bibfnamefont {H.}~\bibnamefont {Weinfurter}}, \bibinfo {author} {\bibfnamefont {D.}~\bibnamefont {Kaszlikowski}}, \bibinfo {author} {\bibfnamefont {W.}~\bibnamefont {Laskowski}},\ and\ \bibinfo {author} {\bibfnamefont {M.}~\bibnamefont {Zukowski}},\ }\bibfield  {title} {\bibinfo {title} {Three-photon w-state},\ }\href@noop {} {\bibfield  {journal} {\bibinfo  {journal} {Journal of Modern Optics}\ }\textbf {\bibinfo {volume} {50}},\ \bibinfo {pages} {1131} (\bibinfo {year} {2003})}\BibitemShut {NoStop}%
\bibitem [{\citenamefont {Tashima}\ \emph {et~al.}(2009)\citenamefont {Tashima}, \citenamefont {Wakatsuki}, \citenamefont {{\"O}zdemir}, \citenamefont {Yamamoto}, \citenamefont {Koashi},\ and\ \citenamefont {Imoto}}]{tashima2009local}%
  \BibitemOpen
  \bibfield  {author} {\bibinfo {author} {\bibfnamefont {T.}~\bibnamefont {Tashima}}, \bibinfo {author} {\bibfnamefont {T.}~\bibnamefont {Wakatsuki}}, \bibinfo {author} {\bibfnamefont {{\c{S}}.~K.}\ \bibnamefont {{\"O}zdemir}}, \bibinfo {author} {\bibfnamefont {T.}~\bibnamefont {Yamamoto}}, \bibinfo {author} {\bibfnamefont {.~f.~M.}\ \bibnamefont {Koashi}},\ and\ \bibinfo {author} {\bibfnamefont {N.}~\bibnamefont {Imoto}},\ }\bibfield  {title} {\bibinfo {title} {Local transformation of two einstein-podolsky-rosen photon pairs into a three-photon w state},\ }\href@noop {} {\bibfield  {journal} {\bibinfo  {journal} {Physical review letters}\ }\textbf {\bibinfo {volume} {102}},\ \bibinfo {pages} {130502} (\bibinfo {year} {2009})}\BibitemShut {NoStop}%
\bibitem [{\citenamefont {Fang}\ \emph {et~al.}(2019)\citenamefont {Fang}, \citenamefont {Menotti}, \citenamefont {Liscidini}, \citenamefont {Sipe},\ and\ \citenamefont {Lorenz}}]{Fang2019PRL}%
  \BibitemOpen
  \bibfield  {author} {\bibinfo {author} {\bibfnamefont {B.}~\bibnamefont {Fang}}, \bibinfo {author} {\bibfnamefont {M.}~\bibnamefont {Menotti}}, \bibinfo {author} {\bibfnamefont {M.}~\bibnamefont {Liscidini}}, \bibinfo {author} {\bibfnamefont {J.~E.}\ \bibnamefont {Sipe}},\ and\ \bibinfo {author} {\bibfnamefont {V.~O.}\ \bibnamefont {Lorenz}},\ }\bibfield  {title} {\bibinfo {title} {Three-photon discrete-energy-entangled $w$ state in an optical fiber},\ }\href {https://doi.org/10.1103/PhysRevLett.123.070508} {\bibfield  {journal} {\bibinfo  {journal} {Phys. Rev. Lett.}\ }\textbf {\bibinfo {volume} {123}},\ \bibinfo {pages} {070508} (\bibinfo {year} {2019})}\BibitemShut {NoStop}%
\bibitem [{\citenamefont {Gr{\"a}fe}\ \emph {et~al.}(2014)\citenamefont {Gr{\"a}fe}, \citenamefont {Heilmann}, \citenamefont {Perez-Leija}, \citenamefont {Keil}, \citenamefont {Dreisow}, \citenamefont {Heinrich}, \citenamefont {Moya-Cessa}, \citenamefont {Nolte}, \citenamefont {Christodoulides},\ and\ \citenamefont {Szameit}}]{grafe2014chip}%
  \BibitemOpen
  \bibfield  {author} {\bibinfo {author} {\bibfnamefont {M.}~\bibnamefont {Gr{\"a}fe}}, \bibinfo {author} {\bibfnamefont {R.}~\bibnamefont {Heilmann}}, \bibinfo {author} {\bibfnamefont {A.}~\bibnamefont {Perez-Leija}}, \bibinfo {author} {\bibfnamefont {R.}~\bibnamefont {Keil}}, \bibinfo {author} {\bibfnamefont {F.}~\bibnamefont {Dreisow}}, \bibinfo {author} {\bibfnamefont {M.}~\bibnamefont {Heinrich}}, \bibinfo {author} {\bibfnamefont {H.}~\bibnamefont {Moya-Cessa}}, \bibinfo {author} {\bibfnamefont {S.}~\bibnamefont {Nolte}}, \bibinfo {author} {\bibfnamefont {D.~N.}\ \bibnamefont {Christodoulides}},\ and\ \bibinfo {author} {\bibfnamefont {A.}~\bibnamefont {Szameit}},\ }\bibfield  {title} {\bibinfo {title} {On-chip generation of high-order single-photon w-states},\ }\href@noop {} {\bibfield  {journal} {\bibinfo  {journal} {Nature Photonics}\ }\textbf {\bibinfo {volume} {8}},\ \bibinfo {pages} {791} (\bibinfo {year} {2014})}\BibitemShut {NoStop}%
\bibitem [{\citenamefont {Horodecki}\ \emph {et~al.}(2009)\citenamefont {Horodecki}, \citenamefont {Horodecki}, \citenamefont {Horodecki},\ and\ \citenamefont {Horodecki}}]{Horodecki2009RMP}%
  \BibitemOpen
  \bibfield  {author} {\bibinfo {author} {\bibfnamefont {R.}~\bibnamefont {Horodecki}}, \bibinfo {author} {\bibfnamefont {P.}~\bibnamefont {Horodecki}}, \bibinfo {author} {\bibfnamefont {M.}~\bibnamefont {Horodecki}},\ and\ \bibinfo {author} {\bibfnamefont {K.}~\bibnamefont {Horodecki}},\ }\bibfield  {title} {\bibinfo {title} {Quantum entanglement},\ }\href {https://doi.org/10.1103/RevModPhys.81.865} {\bibfield  {journal} {\bibinfo  {journal} {Rev. Mod. Phys.}\ }\textbf {\bibinfo {volume} {81}},\ \bibinfo {pages} {865} (\bibinfo {year} {2009})}\BibitemShut {NoStop}%
\bibitem [{\citenamefont {G{\"u}hne}\ and\ \citenamefont {T{\'o}th}(2009)}]{guhne2009PR}%
  \BibitemOpen
  \bibfield  {author} {\bibinfo {author} {\bibfnamefont {O.}~\bibnamefont {G{\"u}hne}}\ and\ \bibinfo {author} {\bibfnamefont {G.}~\bibnamefont {T{\'o}th}},\ }\bibfield  {title} {\bibinfo {title} {Entanglement detection},\ }\href@noop {} {\bibfield  {journal} {\bibinfo  {journal} {Physics Reports}\ }\textbf {\bibinfo {volume} {474}},\ \bibinfo {pages} {1} (\bibinfo {year} {2009})}\BibitemShut {NoStop}%
%
%
\bibitem [{\citenamefont {Chruscinski}\ and\ \citenamefont {Sarbicki}(2014)}]{Chruscinski2014IOP}%
  \BibitemOpen
  \bibfield  {author} {\bibinfo {author} {\bibfnamefont {D.}~\bibnamefont {Chru\'{s}ci\'{n}ski}}\ and\ \bibinfo {author} {\bibfnamefont {G.}~\bibnamefont {Sarbicki}},\ }\bibfield  {title} {\bibinfo {title} {Entanglement witnesses: construction, analysis and classification},\ }\href {https://doi.org/10.1088/1751-8113/47/48/483001} {\bibfield  {journal} {\bibinfo  {journal} {Journal of Physics A: Mathematical and Theoretical}\ }\textbf {\bibinfo {volume} {47}},\ \bibinfo {pages} {483001} (\bibinfo {year} {2014})}\BibitemShut {NoStop}%  
%  
%  
%  
%\bibitem [{\citenamefont {Chruscinski"}\ and\ \citenamefont {Sarbicki}(2014)}]{Chruscinski2014IOP}%
%  \BibitemOpen
%  \bibfield  {author} {\bibinfo {author} {\bibfnamefont {D.}~\bibnamefont {Chru\'{s}ci\'{n}ski}}\ and\ \bibinfo {author} {\bibfnamefont {G.}~\bibnamefont {Sarbicki}},\ }\bibfield  {title} {\bibinfo {title} {Entanglement witnesses: construction, analysis and classification},\ }\href {https://doi.org/10.1088/1751-8113/47/48/483001} {\bibfield  {journal} {\bibinfo  {journal} {Journal of Physics A: Mathematical and Theoretical}\ }\textbf {\bibinfo {volume} {47}},\ \bibinfo {pages} {483001} (\bibinfo {year} {2014})}\BibitemShut {NoStop}%
\bibitem [{\citenamefont {Hillery}\ and\ \citenamefont {Zubairy}(2006)}]{HilleryZubairy2006PRL}%
  \BibitemOpen
  \bibfield  {author} {\bibinfo {author} {\bibfnamefont {M.}~\bibnamefont {Hillery}}\ and\ \bibinfo {author} {\bibfnamefont {M.~S.}\ \bibnamefont {Zubairy}},\ }\bibfield  {title} {\bibinfo {title} {Entanglement conditions for two-mode states},\ }\href {https://doi.org/10.1103/PhysRevLett.96.050503} {\bibfield  {journal} {\bibinfo  {journal} {Phys. Rev. Lett.}\ }\textbf {\bibinfo {volume} {96}},\ \bibinfo {pages} {050503} (\bibinfo {year} {2006})}\BibitemShut {NoStop}%
\bibitem [{\citenamefont {B{\o}rkje}\ \emph {et~al.}(2011)\citenamefont {B{\o}rkje}, \citenamefont {Nunnenkamp},\ and\ \citenamefont {Girvin}}]{Borkje2011PRL}%
  \BibitemOpen
  \bibfield  {author} {\bibinfo {author} {\bibfnamefont {K.}~\bibnamefont {B{\o}rkje}}, \bibinfo {author} {\bibfnamefont {A.}~\bibnamefont {Nunnenkamp}},\ and\ \bibinfo {author} {\bibfnamefont {S.~M.}\ \bibnamefont {Girvin}},\ }\bibfield  {title} {\bibinfo {title} {Proposal for entangling remote micromechanical oscillators via optical measurements},\ }\href {https://doi.org/10.1103/PhysRevLett.107.123601} {\bibfield  {journal} {\bibinfo  {journal} {Phys. Rev. Lett.}\ }\textbf {\bibinfo {volume} {107}},\ \bibinfo {pages} {123601} (\bibinfo {year} {2011})}\BibitemShut {NoStop}%
\bibitem [{\citenamefont {G{\"u}hne}\ and\ \citenamefont {Seevinck}(2010)}]{Guhne2010NJP}%
  \BibitemOpen
  \bibfield  {author} {\bibinfo {author} {\bibfnamefont {O.}~\bibnamefont {G{\"u}hne}}\ and\ \bibinfo {author} {\bibfnamefont {M.}~\bibnamefont {Seevinck}},\ }\bibfield  {title} {\bibinfo {title} {Separability criteria for genuine multiparticle entanglement},\ }\href {https://doi.org/10.1088/1367-2630/12/5/053002} {\bibfield  {journal} {\bibinfo  {journal} {New Journal of Physics}\ }\textbf {\bibinfo {volume} {12}},\ \bibinfo {pages} {053002} (\bibinfo {year} {2010})}\BibitemShut {NoStop}%
\bibitem [{\citenamefont {Aspelmeyer}\ \emph {et~al.}(2014)\citenamefont {Aspelmeyer}, \citenamefont {Kippenberg},\ and\ \citenamefont {Marquardt}}]{Aspelmeyer2014RMP}%
  \BibitemOpen
  \bibfield  {author} {\bibinfo {author} {\bibfnamefont {M.}~\bibnamefont {Aspelmeyer}}, \bibinfo {author} {\bibfnamefont {T.~J.}\ \bibnamefont {Kippenberg}},\ and\ \bibinfo {author} {\bibfnamefont {F.}~\bibnamefont {Marquardt}},\ }\bibfield  {title} {\bibinfo {title} {Cavity optomechanics},\ }\href@noop {} {\bibfield  {journal} {\bibinfo  {journal} {Rev. Mod. Phys.}\ }\textbf {\bibinfo {volume} {86}},\ \bibinfo {pages} {1391} (\bibinfo {year} {2014})}\BibitemShut {NoStop}%
\bibitem [{\citenamefont {Vanner}\ \emph {et~al.}(2013)\citenamefont {Vanner}, \citenamefont {Aspelmeyer},\ and\ \citenamefont {Kim}}]{Vanner2013PRL}%
  \BibitemOpen
  \bibfield  {author} {\bibinfo {author} {\bibfnamefont {M.~R.}\ \bibnamefont {Vanner}}, \bibinfo {author} {\bibfnamefont {M.}~\bibnamefont {Aspelmeyer}},\ and\ \bibinfo {author} {\bibfnamefont {M.~S.}\ \bibnamefont {Kim}},\ }\bibfield  {title} {\bibinfo {title} {Quantum state orthogonalization and a toolset for quantum optomechanical phonon control},\ }\href {https://doi.org/10.1103/PhysRevLett.110.010504} {\bibfield  {journal} {\bibinfo  {journal} {Phys. Rev. Lett.}\ }\textbf {\bibinfo {volume} {110}},\ \bibinfo {pages} {010504} (\bibinfo {year} {2013})}\BibitemShut {NoStop}%
\bibitem [{\citenamefont {Bush}\ and\ \citenamefont {B\o{}rkje}(2024)}]{Bush2024PRA}%
  \BibitemOpen
  \bibfield  {author} {\bibinfo {author} {\bibfnamefont {K.~R.}\ \bibnamefont {Bush}}\ and\ \bibinfo {author} {\bibfnamefont {K.}~\bibnamefont {B\o{}rkje}},\ }\bibfield  {title} {\bibinfo {title} {Proposal for observing nonclassicality in highly excited mechanical oscillators by single photon detection},\ }\href {https://doi.org/10.1103/PhysRevA.109.043505} {\bibfield  {journal} {\bibinfo  {journal} {Phys. Rev. A}\ }\textbf {\bibinfo {volume} {109}},\ \bibinfo {pages} {043505} (\bibinfo {year} {2024})}\BibitemShut {NoStop}%
\bibitem [{\citenamefont {Riedinger}\ \emph {et~al.}(2016)\citenamefont {Riedinger}, \citenamefont {Hong}, \citenamefont {Norte}, \citenamefont {Slater}, \citenamefont {Shang}, \citenamefont {Krause}, \citenamefont {Anant}, \citenamefont {Aspelmeyer},\ and\ \citenamefont {Gr\"{o}blacher}}]{Riedinger2016Nature}%
  \BibitemOpen
  \bibfield  {author} {\bibinfo {author} {\bibfnamefont {R.}~\bibnamefont {Riedinger}}, \bibinfo {author} {\bibfnamefont {S.}~\bibnamefont {Hong}}, \bibinfo {author} {\bibfnamefont {R.~A.}\ \bibnamefont {Norte}}, \bibinfo {author} {\bibfnamefont {J.~A.}\ \bibnamefont {Slater}}, \bibinfo {author} {\bibfnamefont {J.}~\bibnamefont {Shang}}, \bibinfo {author} {\bibfnamefont {A.~G.}\ \bibnamefont {Krause}}, \bibinfo {author} {\bibfnamefont {V.}~\bibnamefont {Anant}}, \bibinfo {author} {\bibfnamefont {M.}~\bibnamefont {Aspelmeyer}},\ and\ \bibinfo {author} {\bibfnamefont {S.}~\bibnamefont {Gr\"{o}blacher}},\ }\bibfield  {title} {\bibinfo {title} {Non-classical correlations between single photons and phonons from a mechanical oscillator},\ }\href@noop {} {\bibfield  {journal} {\bibinfo  {journal} {Nature}\ }\textbf {\bibinfo {volume} {530}},\ \bibinfo {pages} {313} (\bibinfo {year} {2016})}\BibitemShut {NoStop}%
\bibitem [{\citenamefont {Hong}\ \emph {et~al.}(2017)\citenamefont {Hong}, \citenamefont {Riedinger}, \citenamefont {Marinkovi{\'c}}, \citenamefont {Wallucks}, \citenamefont {Hofer}, \citenamefont {Norte}, \citenamefont {Aspelmeyer},\ and\ \citenamefont {Gr{\"o}blacher}}]{Hong2017Science}%
  \BibitemOpen
  \bibfield  {author} {\bibinfo {author} {\bibfnamefont {S.}~\bibnamefont {Hong}}, \bibinfo {author} {\bibfnamefont {R.}~\bibnamefont {Riedinger}}, \bibinfo {author} {\bibfnamefont {I.}~\bibnamefont {Marinkovi{\'c}}}, \bibinfo {author} {\bibfnamefont {A.}~\bibnamefont {Wallucks}}, \bibinfo {author} {\bibfnamefont {S.~G.}\ \bibnamefont {Hofer}}, \bibinfo {author} {\bibfnamefont {R.~A.}\ \bibnamefont {Norte}}, \bibinfo {author} {\bibfnamefont {M.}~\bibnamefont {Aspelmeyer}},\ and\ \bibinfo {author} {\bibfnamefont {S.}~\bibnamefont {Gr{\"o}blacher}},\ }\bibfield  {title} {\bibinfo {title} {Hanbury-{B}rown and {T}wiss interferometry of single phonons from an optomechanical resonator},\ }\href {https://doi.org/10.1126/science.aan7939} {\bibfield  {journal} {\bibinfo  {journal} {Science}\ }\textbf {\bibinfo {volume} {358}},\ \bibinfo {pages} {203} (\bibinfo {year} {2017})}\BibitemShut {NoStop}%
\bibitem [{\citenamefont {Velez}\ \emph {et~al.}(2019)\citenamefont {Velez}, \citenamefont {Seibold}, \citenamefont {Kipfer}, \citenamefont {Anderson}, \citenamefont {Sudhir},\ and\ \citenamefont {Galland}}]{Velez2019PRX}%
  \BibitemOpen
  \bibfield  {author} {\bibinfo {author} {\bibfnamefont {S.~T.}\ \bibnamefont {Velez}}, \bibinfo {author} {\bibfnamefont {K.}~\bibnamefont {Seibold}}, \bibinfo {author} {\bibfnamefont {N.}~\bibnamefont {Kipfer}}, \bibinfo {author} {\bibfnamefont {M.~D.}\ \bibnamefont {Anderson}}, \bibinfo {author} {\bibfnamefont {V.}~\bibnamefont {Sudhir}},\ and\ \bibinfo {author} {\bibfnamefont {C.}~\bibnamefont {Galland}},\ }\bibfield  {title} {\bibinfo {title} {Preparation and decay of a single quantum of vibration at ambient conditions},\ }\href {https://doi.org/10.1103/PhysRevX.9.041007} {\bibfield  {journal} {\bibinfo  {journal} {Phys. Rev. X}\ }\textbf {\bibinfo {volume} {9}},\ \bibinfo {pages} {041007} (\bibinfo {year} {2019})}\BibitemShut {NoStop}%
\bibitem [{\citenamefont {Wang}(2023)}]{Wang2023PhDThesis}%
  \BibitemOpen
  \bibfield  {author} {\bibinfo {author} {\bibfnamefont {Y.}~\bibnamefont {Wang}},\ }\emph {\bibinfo {title} {Manipulating and Measuring States of a Superfluid Optomechanical Resonator in the Quantum Regime}},\ \href {https://harrislab.yale.edu/files/thesis/Wang_Thesis.pdf} {Ph.D. thesis},\ \bibinfo  {school} {Yale University} (\bibinfo {year} {2023})\BibitemShut {NoStop}%
\bibitem [{\citenamefont {Patel}\ \emph {et~al.}(2021)\citenamefont {Patel}, \citenamefont {McKenna}, \citenamefont {Wang}, \citenamefont {Witmer}, \citenamefont {Jiang}, \citenamefont {Van~Laer}, \citenamefont {Sarabalis},\ and\ \citenamefont {Safavi-Naeini}}]{Patel2021PRL}%
  \BibitemOpen
  \bibfield  {author} {\bibinfo {author} {\bibfnamefont {R.~N.}\ \bibnamefont {Patel}}, \bibinfo {author} {\bibfnamefont {T.~P.}\ \bibnamefont {McKenna}}, \bibinfo {author} {\bibfnamefont {Z.}~\bibnamefont {Wang}}, \bibinfo {author} {\bibfnamefont {J.~D.}\ \bibnamefont {Witmer}}, \bibinfo {author} {\bibfnamefont {W.}~\bibnamefont {Jiang}}, \bibinfo {author} {\bibfnamefont {R.}~\bibnamefont {Van~Laer}}, \bibinfo {author} {\bibfnamefont {C.~J.}\ \bibnamefont {Sarabalis}},\ and\ \bibinfo {author} {\bibfnamefont {A.~H.}\ \bibnamefont {Safavi-Naeini}},\ }\bibfield  {title} {\bibinfo {title} {Room-temperature mechanical resonator with a single added or subtracted phonon},\ }\href {https://doi.org/10.1103/PhysRevLett.127.133602} {\bibfield  {journal} {\bibinfo  {journal} {Phys. Rev. Lett.}\ }\textbf {\bibinfo {volume} {127}},\ \bibinfo {pages} {133602} (\bibinfo {year} {2021})}\BibitemShut {NoStop}%
\bibitem [{\citenamefont {Enzian}\ \emph {et~al.}(2021)\citenamefont {Enzian}, \citenamefont {Freisem}, \citenamefont {Price}, \citenamefont {Svela}, \citenamefont {Clarke}, \citenamefont {Shajilal}, \citenamefont {Janousek}, \citenamefont {Buchler}, \citenamefont {Lam},\ and\ \citenamefont {Vanner}}]{Enzian2021non}%
  \BibitemOpen
  \bibfield  {author} {\bibinfo {author} {\bibfnamefont {G.}~\bibnamefont {Enzian}}, \bibinfo {author} {\bibfnamefont {L.}~\bibnamefont {Freisem}}, \bibinfo {author} {\bibfnamefont {J.~J.}\ \bibnamefont {Price}}, \bibinfo {author} {\bibfnamefont {A.~{\O}.}\ \bibnamefont {Svela}}, \bibinfo {author} {\bibfnamefont {J.}~\bibnamefont {Clarke}}, \bibinfo {author} {\bibfnamefont {B.}~\bibnamefont {Shajilal}}, \bibinfo {author} {\bibfnamefont {J.}~\bibnamefont {Janousek}}, \bibinfo {author} {\bibfnamefont {B.~C.}\ \bibnamefont {Buchler}}, \bibinfo {author} {\bibfnamefont {P.~K.}\ \bibnamefont {Lam}},\ and\ \bibinfo {author} {\bibfnamefont {M.~R.}\ \bibnamefont {Vanner}},\ }\bibfield  {title} {\bibinfo {title} {Non-gaussian mechanical motion via single and multiphonon subtraction from a thermal state},\ }\href@noop {} {\bibfield  {journal} {\bibinfo  {journal} {Physical Review Letters}\ }\textbf {\bibinfo {volume} {127}},\ \bibinfo {pages} {243601} (\bibinfo {year} {2021})}\BibitemShut {NoStop}%
\bibitem [{\citenamefont {Reck}\ \emph {et~al.}(1994)\citenamefont {Reck}, \citenamefont {Zeilinger}, \citenamefont {Bernstein},\ and\ \citenamefont {Bertani}}]{Reck1994PRL}%
  \BibitemOpen
  \bibfield  {author} {\bibinfo {author} {\bibfnamefont {M.}~\bibnamefont {Reck}}, \bibinfo {author} {\bibfnamefont {A.}~\bibnamefont {Zeilinger}}, \bibinfo {author} {\bibfnamefont {H.~J.}\ \bibnamefont {Bernstein}},\ and\ \bibinfo {author} {\bibfnamefont {P.}~\bibnamefont {Bertani}},\ }\bibfield  {title} {\bibinfo {title} {Experimental realization of any discrete unitary operator},\ }\href {https://doi.org/10.1103/PhysRevLett.73.58} {\bibfield  {journal} {\bibinfo  {journal} {Phys. Rev. Lett.}\ }\textbf {\bibinfo {volume} {73}},\ \bibinfo {pages} {58} (\bibinfo {year} {1994})}\BibitemShut {NoStop}%
\bibitem [{\citenamefont {Clements}\ \emph {et~al.}(2016)\citenamefont {Clements}, \citenamefont {Humphreys}, \citenamefont {Metcalf}, \citenamefont {Kolthammer},\ and\ \citenamefont {Walmsley}}]{Clements2016optica}%
  \BibitemOpen
  \bibfield  {author} {\bibinfo {author} {\bibfnamefont {W.~R.}\ \bibnamefont {Clements}}, \bibinfo {author} {\bibfnamefont {P.~C.}\ \bibnamefont {Humphreys}}, \bibinfo {author} {\bibfnamefont {B.~J.}\ \bibnamefont {Metcalf}}, \bibinfo {author} {\bibfnamefont {W.~S.}\ \bibnamefont {Kolthammer}},\ and\ \bibinfo {author} {\bibfnamefont {I.~A.}\ \bibnamefont {Walmsley}},\ }\bibfield  {title} {\bibinfo {title} {Optimal design for universal multiport interferometers},\ }\href {https://doi.org/10.1364/OPTICA.3.001460} {\bibfield  {journal} {\bibinfo  {journal} {Optica}\ }\textbf {\bibinfo {volume} {3}},\ \bibinfo {pages} {1460} (\bibinfo {year} {2016})}\BibitemShut {NoStop}%
\bibitem [{\citenamefont {Meenehan}\ \emph {et~al.}(2015)\citenamefont {Meenehan}, \citenamefont {Cohen}, \citenamefont {MacCabe}, \citenamefont {Marsili}, \citenamefont {Shaw},\ and\ \citenamefont {Painter}}]{Meenehan2015PRX}%
  \BibitemOpen
  \bibfield  {author} {\bibinfo {author} {\bibfnamefont {S.~M.}\ \bibnamefont {Meenehan}}, \bibinfo {author} {\bibfnamefont {J.~D.}\ \bibnamefont {Cohen}}, \bibinfo {author} {\bibfnamefont {G.~S.}\ \bibnamefont {MacCabe}}, \bibinfo {author} {\bibfnamefont {F.}~\bibnamefont {Marsili}}, \bibinfo {author} {\bibfnamefont {M.~D.}\ \bibnamefont {Shaw}},\ and\ \bibinfo {author} {\bibfnamefont {O.}~\bibnamefont {Painter}},\ }\bibfield  {title} {\bibinfo {title} {Pulsed excitation dynamics of an optomechanical crystal resonator near its quantum ground state of motion},\ }\href {https://doi.org/10.1103/PhysRevX.5.041002} {\bibfield  {journal} {\bibinfo  {journal} {Phys. Rev. X}\ }\textbf {\bibinfo {volume} {5}},\ \bibinfo {pages} {041002} (\bibinfo {year} {2015})}\BibitemShut {NoStop}%
\bibitem [{\citenamefont {Galinskiy}\ \emph {et~al.}(2020)\citenamefont {Galinskiy}, \citenamefont {Tsaturyan}, \citenamefont {Parniak},\ and\ \citenamefont {Polzik}}]{Galinskiy2020Optica}%
  \BibitemOpen
  \bibfield  {author} {\bibinfo {author} {\bibfnamefont {I.}~\bibnamefont {Galinskiy}}, \bibinfo {author} {\bibfnamefont {Y.}~\bibnamefont {Tsaturyan}}, \bibinfo {author} {\bibfnamefont {M.}~\bibnamefont {Parniak}},\ and\ \bibinfo {author} {\bibfnamefont {E.~S.}\ \bibnamefont {Polzik}},\ }\bibfield  {title} {\bibinfo {title} {Phonon counting thermometry of an ultracoherent membrane resonator near its motional ground state},\ }\href {https://doi.org/10.1364/OPTICA.390939} {\bibfield  {journal} {\bibinfo  {journal} {Optica}\ }\textbf {\bibinfo {volume} {7}},\ \bibinfo {pages} {718} (\bibinfo {year} {2020})}\BibitemShut {NoStop}%
\bibitem [{\citenamefont {Patil}\ \emph {et~al.}(2022)\citenamefont {Patil}, \citenamefont {Yu}, \citenamefont {Frazier}, \citenamefont {Wang}, \citenamefont {Johnson}, \citenamefont {Fox}, \citenamefont {Reichel},\ and\ \citenamefont {Harris}}]{Patil2022PRL}%
  \BibitemOpen
  \bibfield  {author} {\bibinfo {author} {\bibfnamefont {Y.~S.~S.}\ \bibnamefont {Patil}}, \bibinfo {author} {\bibfnamefont {J.}~\bibnamefont {Yu}}, \bibinfo {author} {\bibfnamefont {S.}~\bibnamefont {Frazier}}, \bibinfo {author} {\bibfnamefont {Y.}~\bibnamefont {Wang}}, \bibinfo {author} {\bibfnamefont {K.}~\bibnamefont {Johnson}}, \bibinfo {author} {\bibfnamefont {J.}~\bibnamefont {Fox}}, \bibinfo {author} {\bibfnamefont {J.}~\bibnamefont {Reichel}},\ and\ \bibinfo {author} {\bibfnamefont {J.~G.~E.}\ \bibnamefont {Harris}},\ }\bibfield  {title} {\bibinfo {title} {Measuring high-order phonon correlations in an optomechanical resonator},\ }\href {https://doi.org/10.1103/PhysRevLett.128.183601} {\bibfield  {journal} {\bibinfo  {journal} {Phys. Rev. Lett.}\ }\textbf {\bibinfo {volume} {128}},\ \bibinfo {pages} {183601} (\bibinfo {year} {2022})}\BibitemShut {NoStop}%
\bibitem [{\citenamefont {G{\"u}hne}\ \emph {et~al.}(2007)\citenamefont {G{\"u}hne}, \citenamefont {Lu}, \citenamefont {Gao},\ and\ \citenamefont {Pan}}]{guhne2007toolbox}%
  \BibitemOpen
  \bibfield  {author} {\bibinfo {author} {\bibfnamefont {O.}~\bibnamefont {G{\"u}hne}}, \bibinfo {author} {\bibfnamefont {C.-Y.}\ \bibnamefont {Lu}}, \bibinfo {author} {\bibfnamefont {W.-B.}\ \bibnamefont {Gao}},\ and\ \bibinfo {author} {\bibfnamefont {J.-W.}\ \bibnamefont {Pan}},\ }\bibfield  {title} {\bibinfo {title} {Toolbox for entanglement detection and fidelity estimation},\ }\href@noop {} {\bibfield  {journal} {\bibinfo  {journal} {Physical Review A—Atomic, Molecular, and Optical Physics}\ }\textbf {\bibinfo {volume} {76}},\ \bibinfo {pages} {030305} (\bibinfo {year} {2007})}\BibitemShut {NoStop}%
\bibitem [{\citenamefont {Zhao}\ \emph {et~al.}(2019)\citenamefont {Zhao}, \citenamefont {Wang}, \citenamefont {Yuan},\ and\ \citenamefont {Ma}}]{zhao2019efficient}%
  \BibitemOpen
  \bibfield  {author} {\bibinfo {author} {\bibfnamefont {Q.}~\bibnamefont {Zhao}}, \bibinfo {author} {\bibfnamefont {G.}~\bibnamefont {Wang}}, \bibinfo {author} {\bibfnamefont {X.}~\bibnamefont {Yuan}},\ and\ \bibinfo {author} {\bibfnamefont {X.}~\bibnamefont {Ma}},\ }\bibfield  {title} {\bibinfo {title} {Efficient and robust detection of multipartite greenberger-horne-zeilinger-like states},\ }\href@noop {} {\bibfield  {journal} {\bibinfo  {journal} {Physical Review A}\ }\textbf {\bibinfo {volume} {99}},\ \bibinfo {pages} {052349} (\bibinfo {year} {2019})}\BibitemShut {NoStop}%
\bibitem [{\citenamefont {Zhou}(2020)}]{zhou2020entanglement}%
  \BibitemOpen
  \bibfield  {author} {\bibinfo {author} {\bibfnamefont {Y.}~\bibnamefont {Zhou}},\ }\bibfield  {title} {\bibinfo {title} {Entanglement detection under coherent noise: Greenberger-horne-zeilinger-like states},\ }\href@noop {} {\bibfield  {journal} {\bibinfo  {journal} {Physical Review A}\ }\textbf {\bibinfo {volume} {101}},\ \bibinfo {pages} {012301} (\bibinfo {year} {2020})}\BibitemShut {NoStop}%
\bibitem [{\citenamefont {Landon-Cardinal}\ \emph {et~al.}(2018)\citenamefont {Landon-Cardinal}, \citenamefont {Govia},\ and\ \citenamefont {Clerk}}]{landon2018quantitative}%
  \BibitemOpen
  \bibfield  {author} {\bibinfo {author} {\bibfnamefont {O.}~\bibnamefont {Landon-Cardinal}}, \bibinfo {author} {\bibfnamefont {L.~C.}\ \bibnamefont {Govia}},\ and\ \bibinfo {author} {\bibfnamefont {A.~A.}\ \bibnamefont {Clerk}},\ }\bibfield  {title} {\bibinfo {title} {Quantitative tomography for continuous variable quantum systems},\ }\href@noop {} {\bibfield  {journal} {\bibinfo  {journal} {Physical review letters}\ }\textbf {\bibinfo {volume} {120}},\ \bibinfo {pages} {090501} (\bibinfo {year} {2018})}\BibitemShut {NoStop}%
\end{thebibliography}
%\input{main.bbl}


\newpage

\appendix
\section{Expressing \texorpdfstring{$N_{\text{sep.},\mathcal{C}, \text{max}}$}{Nsep.,C,max} by $N$ and $M$} \label{app:Finding_NsepCmax}
A central quantity in the bound \eqref{eq:absNMineq} is the maximal number of unique separable pairs $N_{\text{sep.},\mathcal{C}, \text{max}}$ a structure $S=\{ C_k\}$ can exhibit. If one considers every possible structure, this value is trivially found to be $N(N-1)/2$, corresponding to $N$ combinations of size $1$. However, under the constraint of a maximal combination size $M$ and that no two combinations can have a combined size $M$ or less, finding this maximum becomes nontrivial. Here we show, using a combination of analytical and algebraic methods, that the solution can always be found and written in closed form.

In general, among structures consisting of a given number $\mathcal{N}_C$ of combinations, the number of separable pairs is bounded by
\begin{equation} \label{eq:Nsep_upperbound}
\begin{aligned}
    N_\text{sep.} = \sum_{k=1}^{\mathcal{N}_C-1} \sum_{l=k+1}^{\mathcal{N}_C} \abs{C_k} \abs{C_l} &= \frac{N^2-\sum_{k=1}^{\mathcal{N}_C} \abs{C_k}^2}{2} \\
    &\leq \frac{N^2-\mathcal{N}_C \left( \sum_{k=1}^{\mathcal{N}_C} \frac{\abs{C_k}}{\mathcal{N}_C} \right)^2}{2} \\
    &= \frac{N^2}{2}\left(1-\frac{1}{\mathcal{N}_C}\right),
\end{aligned}
\end{equation}
where the middle line is obtained through convexity of the square. Equality is obtained when all combinations are of identical size $\abs{C_k} = N/\mathcal{N}_C$. If $N/\mathcal{N}_C$ is not an integer, it is still useful to write the true, discrete, maximum as $\abs{C_k} = N/\mathcal{N}_C + r_k$, which deviates from the upper bound in \eqref{eq:Nsep_upperbound} by
\begin{equation}
\begin{aligned}
    &\abs{\frac{N^2}{2}\left(1-\frac{1}{\mathcal{N}_C}\right)-\frac{N^2-\sum_{k=1}^{\mathcal{N}_C} (N/\mathcal{N}_C + r_k)^2}{2}} \\
    &= 
    \abs{\frac{1}{2}\left( \frac{2N}{\mathcal{N}_C} \sum_{k=1}^{\mathcal{N}_C} r_k + \sum_{k=1}^{\mathcal{N}_C} r_k^2 \right)} = \frac{1}{2} \sum_{k=1}^{\mathcal{N}_C} r_k^2,
\end{aligned}
\end{equation}
where we use that $\sum_{k=1}^{\mathcal{N}_C} r_k = 0$. Thus the maximum is obtained when the euclidean distance between the vectors $(\abs{C_1}, \abs{C_2},\ldots)$ and $N/\mathcal{N}_C(1, 1,\ldots)$ in $\reals^{\mathcal{N}_C}$ is minimized, under the constraint $\sum_{k=1}^{\mathcal{N}_C} \abs{C_k} = N$. Geometrically, considering a plot in $\reals^{\mathcal{N}_C}$ in which integer increments are indicated, this means that the solution is one of the corners of the ``grid cell" in which the point $N/\mathcal{N}_C(1, 1,\ldots)$ lies. The solution is found by populating $\mathcal{N}_C$ combinations with $\lfloor N/\mathcal{N}_C \rfloor$ modes each, and then distributing the remaining $N-\mathcal{N}_C\lfloor N/\mathcal{N}_C \rfloor$ modes as evenly as possible between them. The floor function $\lfloor \cdot \rfloor$ indicates rounding down to the nearest integer, such that for integers $A \geq B$, $\lfloor A/B \rfloor$ is the result of integer division, and $A-\lfloor A/B \rfloor B$ is the remainder.

Our remaining degree of freedom is $\mathcal{N}_C$ itself. In order to find $N_{\text{sep.},\mathcal{C}, \text{max}}$, it must be chosen such that it is as large as possible without violating our assumption of maximally $M$ modes per combination, or the assumption that no two combinations have combined size less than or equal to $M$. 

For odd values of $M$, this is always satisfied by choosing 
\begin{equation}
    \mathcal{N}_C = \bigg\lfloor\frac{2N}{M+1}\bigg\rfloor,
\end{equation}
which corresponds to how many combinations of size $(M+1)/2$ fit into a set of $N$ elements. With all combinations being of size $(M+1)/2$ or greater, the corresponding structure is guaranteed to be irreducible. The choice $\mathcal{N}_C = \lfloor 2N/(M+1) \rfloor+1$, on the other hand, would lead to combinations of both sizes $(M+1)/2$ and $(M+1)/2-1$, thus the result would always be reducible.

For even values of $M$, the choice $\mathcal{N}_C = \lfloor 2N/M \rfloor$ leads to an irreducible structure only if the remainder is $N-\mathcal{N}_C M/2 = \mathcal{N}_C-1$. This can be understood by first dividing the modes into combinations of size $M/2$, and then allocating the remaining modes as evenly as possible between those combinations. If the remainder is less than $\mathcal{N}_C-1$, there will be at least two combinations whose sizes are $M/2$, making the structure reducible. Otherwise, all pairs of combinations will have a combined size of at least $M+1$. If this is not satisfied by $\mathcal{N}_C = \lfloor 2N/M \rfloor$, we may reduce the number of combinations by $1$ and try again, repeating this process $A$ times until the remainder satisfies $N-(\lfloor 2N/M \rfloor-A)M/2 \geq (\lfloor 2N/M \rfloor-A)-1$, meaning that either the original criterion is satisfied, or all combinations are of at least size $M/2+1$. This condition permits solving for $A$, and we find that
\begin{equation}
    A = \lfloor 2N/M \rfloor - \bigg\lfloor\frac{2(N+1)}{M+2}\bigg\rfloor
\end{equation}
is the first integer solution to satisfy the constraint. This produces the simple solution for the optimal number of combinations,
\begin{equation}
    \mathcal{N}_C = \bigg\lfloor\frac{2(N+1)}{M+2}\bigg\rfloor,
\end{equation}
for even $M$.

Lastly, we note that if the optimal structure for any given choice of $\mathcal{N}_C$ is reducible, all other structures consisting of the same number of combinations $\mathcal{N}_C$ are also reducible; any increase in size of one combination requires the decrease of another due to constancy of $\sum_{k=1}^{\mathcal{N}_C} \abs{C_k} = N$, which will always lead to the same, or some other, pair of combinations having combined size less than or equal to $M$. Thus our maximization with regards to $\mathcal{N}_C$ above produces the structures with the maximal possible number of separable pairs among all irreducible structures.

Our solutions consist of $N-\lfloor N/\mathcal{N}_C \rfloor \mathcal{N}_C$ combinations of size $\lfloor N/\mathcal{N}_C \rfloor+1$ and $\mathcal{N}_C-(N-\lfloor N/\mathcal{N}_C \rfloor \mathcal{N}_C)$ combinations of size $\lfloor N/\mathcal{N}_C \rfloor$. Using these expressions on the right hand side of the first line in \eqref{eq:Nsep_upperbound}, we finally obtain
\begin{equation}
    N_{\text{sep.},\mathcal{C}, \text{max}} =
        \frac{N(N-1)-(2N-\mathcal{N}_C) \big\lfloor\frac{N}{\mathcal{N}_C}\big\rfloor+ \mathcal{N}_C \big\lfloor\frac{N}{\mathcal{N}_C}\big\rfloor^2}{2},
\end{equation}
where
\begin{equation}
    \mathcal{N}_C = \begin{cases}
        &\big\lfloor\frac{2(N+1)}{M+2}\big\rfloor, \ \ \ M \ \text{even} \\
        &\big\lfloor\frac{2N}{M+1}\big\rfloor, \ \ \ \ \ \ M \ \text{odd}.
    \end{cases}
\end{equation}

\section{Optimal choices of \texorpdfstring{$N$}{N} for detecting \texorpdfstring{$M+1$}{M+1}-partite entanglement and nonlocality} \label{app:Opt_N}
A curious feature of our witnesses as applied to thermal $W$-like states \eqref{eq:threshold_sym_thermal}, \eqref{eq:nonlocbound_distinguishM_thermal} is that the thermal thresholds for violating the inequalities are not monotonously decreasing functions of $N$. It turns out that for every degree $M+1$ of genuine entanglement one might wish to witness, there exists some $N = N_\text{opt} \geq M+1$ which maximizes the threshold thermal occupation number below which the witness is violated. 

\begin{figure}[]
	% \begin{tikzpicture}[scale=1]
	% \begin{axis}[xlabel=$M+1$, ylabel={$N_\text{opt.}$}, grid=both, ymax=80, ymin=0, xmin=0, xmax=40]
	% \addplot[color=red] table [x index=0,y index=2, col sep=comma] {Plots/thresh_abswitness_trick_optN.csv};
 %    \addplot[color=blue] table [x index=0,y index=2, col sep=comma] {Plots/thresh_relwitness_max_optN.csv};
	% \end{axis}
	% \end{tikzpicture}
    \includegraphics{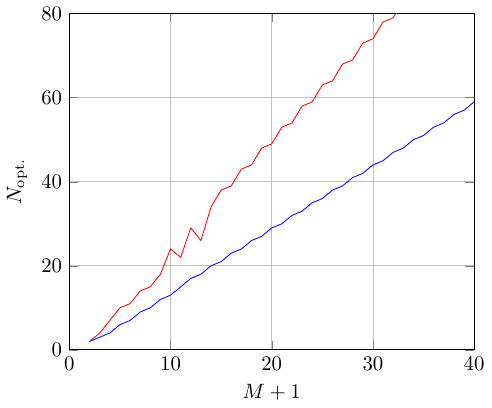}
	\caption{\label{fig:optN}
        The optimal value $N_\text{opt.}$ for which $M+1$-partite entanglement can most easily be measured. The values for the entanglement witness \eqref{eq:threshold_sym_thermal} (in red) is approximately linear with a slope of $\sim 2.5$. The fluctuations are combinatoric in nature, with sharp variations around prime numbered values of $M+1$. The case of the nonlocality witness \eqref{eq:nonlocbound_distinguishM_thermal} (in blue) is smoother, and with a slope of $\sim 1.6$.
		}
\end{figure}

By optimizing the thresholds numerically over values of $N \geq M+1$, we obtain the results shown in \fref{fig:achievableMplus1_Nopt_entwitness}, \fref{fig:achievableMplus1_Nopt_nonlocwitness} and \fref{fig:achievableN} (rightmost), constituting a significant advantage compared to applying the witnesses to a fixed number $N=M+1$ of modes. This optimization identifies the optimal number of subsystems $N_\text{opt.}$ for each value of $M+1$. As shown in \fref{fig:optN}, these optimal values increase roughly linearly with $M+1$, with slopes $\sim 2.5$ for the entanglement witness, and $\sim 1.6$ for the nonlocality witness. Meanwhile, the advantage is nonlinear, such that, e.g., increasing $N$ by $\sim 2.5$ times can lead to the same degree of entanglement being witnessable at significantly more than $2.5$ times the thermal occupation.

\onecolumngrid
\end{document}